\documentclass[10pt,a4paper]{article}

\usepackage[margin=25mm]{geometry}
\usepackage[T1]{fontenc}
\usepackage[utf8]{inputenc}

\usepackage{amsmath,mathtools,amsthm}
\usepackage{newpxtext}

\usepackage{newpxmath}
\usepackage{microtype}

\usepackage[notrig]{physics}
\RenewDocumentCommand\abs{s m}{\IfBooleanTF{#1}
  {\lvert#2\rvert}
  {\mathopen{}\mathclose\bgroup\left\lvert#2\aftergroup\egroup\right\rvert}}
\RenewDocumentCommand\norm{s m}{\IfBooleanTF{#1}
  {\lVert#2\rVert}
  {\mathopen{}\mathclose\bgroup\left\lVert#2\aftergroup\egroup\right\rVert}}
\usepackage{booktabs}
\usepackage{graphicx,tikz}
\usepackage{enumitem}
\usepackage[numbers]{natbib}
\usepackage{authblk}

\usepackage{xcolor}
\definecolor{linkblue}{RGB}{0,0,139}

\usepackage[
    colorlinks=true,
    linkcolor=linkblue,
    citecolor=linkblue,
    urlcolor=linkblue
]{hyperref}
\hypersetup{
    pdftitle={Quantum resources constrain function approximation},
    pdfauthor={Oxana Shaya, Lirand\"e Pira}
}

\setlist[enumerate]{leftmargin=*,nosep,topsep=3pt}

\newcommand{\ii}{\mathrm i}
\newcommand{\id}{\mathbf 1}
\newcommand{\tp}{\mathsf T}
\newcommand{\params}{\boldsymbol\theta}
\newcommand{\X}{\mathcal X}
\newcommand{\cont}{\mathcal C}
\newcommand{\N}{\mathcal N}
\newcommand{\odd}{\mathrm{odd}}
\DeclareMathOperator{\dist}{dist}
\DeclareMathOperator{\osc}{osc}
\newcommand{\supnorm}[1]{\norm*{#1}_\infty}
\newcommand{\opnorm}[1]{\norm*{#1}_{\mathrm{op}}}
\newcommand{\enorm}[1]{\norm*{#1}_2}

\newtheorem{theorem}{Theorem}
\newtheorem{lemma}{Lemma}[section]
\newtheorem{proposition}[lemma]{Proposition}
\newtheorem*{recalledstatement}{\recalledheading}
\newenvironment{restatement}[2]
  {\begin{samepage}\def\recalledheading{#1~\ref{#2} (restated)}\begin{recalledstatement}}
  {\end{recalledstatement}\end{samepage}}
\AddToHook{env/theorem/begin}{\interlinepenalty=10000 \postdisplaypenalty=10000 }
\AddToHook{env/lemma/begin}{\interlinepenalty=10000 \postdisplaypenalty=10000 }

\title{\textbf{Quantum resources constrain function approximation}}

\author[1]{Oxana Shaya}
\author[2]{Lirand\"e Pira}

\affil[1]{Institute for Information Processing (tnt/L3S), Leibniz Universit\"at Hannover, Hannover, Germany}

\affil[2]{Centre for Quantum Technologies,
National University of Singapore, Singapore}

\date{November 23, 2026}

\begin{document}
\maketitle

\begin{abstract}
\noindent
Universal function approximation is central to machine learning. Quantum learning models are universal as well, but this does not specify the resources needed to approximate a given function. We study how coherence, imaginarity, entanglement and magic constrain function approximation for a fixed encoding and readout, using the cumulative resource power of the trainable gates. In a one-qubit model with phase encoding, coherence bounds the oscillation of the model function, while imaginarity bounds its odd part. In a two-qubit model, entanglement bounds the distance from product functions. These bounds hold at any encoding depth.
For magic, the phase encoding itself supplies non-Clifford operations. On domains bounded away from zero, Clifford trainable gates suffice for uniform approximation. Replacing unrestricted trainable gates by Clifford gates increases the encoding depth at most polynomially in the original depth, the inverse error and the inverse distance of the domain from zero. Magic in the trainable gates can nevertheless improve accuracy at fixed depth. These results show how the resources of a quantum model must match the structure of the target function.
\end{abstract}

\tableofcontents

\section{Introduction}
\label{sec:intro}

A classical neuron applies an activation function to an affine function of its input~\cite[Sec.~6.3]{Goodfellow2016}, an idea that goes back to McCulloch and Pitts~\cite{McCulloch1943} and Rosenblatt~\cite{Rosenblatt1958}.
The activation must not itself be affine, because compositions of affine maps are affine.
With any continuous activation that is not a polynomial, however, neural networks approximate every continuous function on a compact set~\cite{Leshno1993}.
This approximation capacity underlies the use of neural networks as models of nonlinear input--output relations~\cite[Sec.~6.4.1]{Goodfellow2016}.

Quantum learning models obtain their nonlinearity in a different way.
Unitary evolution is linear in the state, but the map from the input $x$ to the encoded state is not~\cite{Mitarai2018,SchuldKilloran2019}.
A data re-uploading circuit encodes the input $x$ several times, with trainable gates in between, and returns an expectation value $f(x)$ as its output~\cite{PerezSalinas2020}; we call $f$ the model function.
For encodings by Pauli rotations with angle $x$, $f$ is a trigonometric polynomial in $x$~\cite{Schuld2021}, and a single qubit already suffices for universal approximation~\cite{PerezSalinas2021,Yu2022}.
Quantum variants of Kolmogorov--Arnold networks use such one-qubit circuits as trainable activation functions~\cite{Jiang2025} or implement the activations by quantum singular value transformation (QSVT) of block-encoded inputs~\cite{Ivashkov_2026}.

Universal approximation establishes the expressive capacity of a model, but does not determine the resources required to approximate a given target function. Previous work has investigated the accessible frequency spectra~\cite{Schuld2021}, the depth requirements for fixed encoding frequencies~\cite{LiuPira2026}, and the classical representation of quantum model functions~\cite{MasotLlima2025}. We ask how individual quantum resources constrain the functions that a model can approximate and derive lower bounds on the resources required for a given target.

We fix the encoding gate and the measured observable. Each trainable gate is assigned its resource power, the largest increase of a resource measure that the gate can cause. The sum over the trainable gates is the cumulative resource power. The encoding and measurement are not included in this sum, even when they can generate resourceful states.

\enlargethispage{\baselineskip}
A simple example shows why the encoding matters.
Consider one qubit that starts in $\ket{0}$ and is measured with $Z$, and the target $g(x)=\cos4x$ on $[-1,1]$.
If the input enters through the rotation $R_y(x)$, four encoding calls realize $g$ with identity trainable gates, because this gate itself creates coherence between the basis states.
If it enters through the phase gate $R_z(x)$, the trainable gates must create this coherence.
Realizing $g$ then requires cumulative cohering power at least $2$ at any depth, and two trainable rotations attain this value (Section~\ref{sec:coherence}).
A resource requirement is a property of the model, with its encoding and readout, and not of the target alone.
\newpage
\paragraph{Main contributions.}
\begin{enumerate}[leftmargin=*,itemsep=4pt,topsep=3pt,beginpenalty=10000]
 \item \textbf{Coherence and imaginarity.}
 A state has coherence if it is not diagonal in the computational basis, and imaginarity if its density matrix is not real.
 For phase encoding and $Z$ readout, cumulative cohering power bounds the oscillation of the model function at any depth and its nonaffinity at fixed depth, while cumulative imaginarity power bounds its odd part (Theorems~\ref{thm:c1} and~\ref{thm:i1}).
 In phase encoding the input is the angle of a rotation about the $z$ axis.
 The oscillation is the difference between the largest and the smallest value of the function.
 The nonaffinity is the sum of the absolute values of its Chebyshev coefficients beyond degree one; it vanishes exactly for affine functions and bounds the distance from them.
 The depth is the number of layers of the circuit, each of which encodes the input once and applies one trainable gate.
 With zero cohering power only constant model functions remain. With zero imaginarity power exactly the continuous even targets can be approximated uniformly.

 \item \textbf{Two-qubit entanglement.}
 With each input encoded locally on its own qubit, trainable two-qubit gates $R_{ZZ}$ and $Z\otimes Z$ readout, cumulative entangling power bounds the distance from product functions, which have one factor for each input, at any depth. At zero power, exactly the bounded product functions are uniformly approximable (Theorem~\ref{thm:entangling-power}). Without a restriction on the entangling power, the model approximates every continuous target with values in $[-1,1]$ uniformly (Appendix~\ref{app:two-qubit-universality}).

 \item \textbf{Approximation with Clifford trainable gates.}
 Magic is a resource for universal quantum computation beyond stabilizer operations~\cite{BravyiKitaev2005}, and Clifford gates are the gates that cannot create it. In our model with phase encoding, the encoding gate already supplies non-Clifford operations for every input except $x=0$, where it is the identity. On domains bounded away from this input, Clifford trainable gates therefore suffice for uniform approximation. Replacing unrestricted trainable gates by Clifford gates increases the depth at most polynomially in the original depth, the inverse error and the inverse distance of the domain from $x=0$ (Theorem~\ref{thm:magic-mean}).
\end{enumerate}
Table~\ref{tab:resource-summary} summarizes the resulting coherence, imaginarity and entanglement requirements.

\begin{table}[t]
\centering
\caption{Resource requirements that hold at any depth.
For each resource, the cumulative power $P_R$ of the trainable gates~\eqref{eq:resource-budget} bounds one property of the model function $f$: its oscillation $\osc(f)$, its odd part $f_{\odd}$, or its uniform distance~\eqref{eq:dist} from the bounded product functions $\mathcal P$~\eqref{eq:entangling-budget}.
All three bounds are sharp.
The last column gives the power that is necessary, but not sufficient, to approximate a target $g$ with uniform error $\varepsilon$, where $(a)_+=\max\{a,0\}$.}
\label{tab:resource-summary}
\small
\setlength{\tabcolsep}{2.5pt}
\renewcommand{\arraystretch}{1.4}
\begin{tabular*}{\linewidth}{@{\extracolsep{\fill}}lclll@{}}
\toprule
Resource & Thm. & \shortstack[l]{Uniform limits\\at $P_R=0$} & Bound on $f$ & \shortstack[l]{Required $P_R$ for\\$\supnorm{f-g}\le\varepsilon$}\\
\midrule
\multicolumn{5}{@{}l}{\emph{One qubit in $\rho_r=(\id+rZ)/2$, phase encoding $R_z(x)$ with $x\in[-1,1]$, $Z$ readout}}\\
Coherence & \ref{thm:c1} & constants $f\equiv\pm r$
 & $\osc(f)\le2r\min\{1,P_C^2/4\}$
 & $P_C\ge\sqrt{2(\osc(g)-2\varepsilon)_+/r}$\\
Imaginarity & \ref{thm:i1} & even functions $f(-x)=f(x)$
 & $\supnorm{f_{\odd}}\le r\min\{1,P_I\}$
 & $P_I\ge(\supnorm{g_{\odd}}-\varepsilon)_+/r$\\
\addlinespace
\multicolumn{5}{@{}l}{\emph{Two qubits in $\ket{00}$, each encoding its own input, $R_{ZZ}$ gates, $Z\otimes Z$ readout}}\\
Entanglement & \ref{thm:entangling-power} & products $f(x_1,x_2)=a(x_1)b(x_2)$
 & $\dist(f,\mathcal P)\le\min\{1,P_E\}$
 & $P_E\ge(\dist(g,\mathcal P)-\varepsilon)_+$\\
\bottomrule
\end{tabular*}
\end{table}

\paragraph{Outline.}
Section~\ref{sec:preliminaries} introduces the resource measures, the power of a gate and the notions of approximation.
Section~\ref{sec:model} defines the one-qubit model with its three encodings and states its universality.
Section~\ref{sec:local} gives the coherence and imaginarity bounds.
Section~\ref{sec:multi} introduces the two-qubit model and relates entangling power to the distance from product functions.
Section~\ref{sec:magic} shows that Clifford trainable gates suffice for approximation on domains bounded away from zero.
Section~\ref{sec:programme} summarizes the results and discusses their dependence on the model, the connection to generative learning, and open questions on resources, training and generalization.
The proofs are collected in the appendices.

\section{Framework and notation}
\label{sec:preliminaries}
We first define the resource measures, the power of a trainable gate and the approximation error.
A state $\rho$ is a positive semidefinite matrix of trace one, and an observable $M$ is Hermitian. We write $\id$ for the identity. For Hermitian $A$ and $B$, the order $A\le B$ means that $B-A$ has nonnegative eigenvalues.
Matrix entries, the transpose $A^{\tp}$ and the entrywise conjugate $\overline A$ refer to the computational basis. The adjoint is $A^\dagger=(\overline A)^{\tp}$.
With $\enorm{v}$ denoting the Euclidean norm, the operator norm is $\opnorm{A}=\max_{\enorm{v}=1}\enorm{Av}$. The trace norm is $\norm*{A}_1=\Tr{\sqrt{A^\dagger A}}$.

\subsection{Resource measures}
\label{sec:measures}
A resource theory describes what can be done under a restriction~\cite{ChitambarGour2019}.
It specifies free states, which are available at no cost, and free operations, which cannot turn a free state into a state that is not free.
All other states are resourceful. A resource measure quantifies their resource content: it vanishes on the free states and cannot increase under free operations~\cite[Sec.~VI.A]{ChitambarGour2019}.
We consider coherence, imaginarity and magic for one qubit, and entanglement for two qubits.
In each case the free states form a closed convex set $\mathcal F$ that contains the maximally mixed state.
We measure each resource by its generalized robustness, called global robustness in Ref.~\cite[Sec.~VI.E.2]{ChitambarGour2019}. It is the least $t\ge0$ such that adding $t$ times some state $\tau$ to $\rho$ and normalizing gives a free state,
\begin{equation}
 R_{\mathcal F}(\rho)=\min\left\{t\ge0:\ \frac{\rho+t\tau}{1+t}\in\mathcal F
 \text{ for some state }\tau\right\}.
\label{eq:robustness}
\end{equation}
It vanishes exactly on $\mathcal F$.
With $\sigma$ denoting the free state $(\rho+t\tau)/(1+t)$, the condition in~\eqref{eq:robustness} reads $(1+t)\sigma-\rho=t\tau\ge0$.
The robustness is therefore also the least $t\ge0$ with $\rho\le(1+t)\sigma$ for some $\sigma\in\mathcal F$.
A quantum operation $\Phi$ that maps free states to free states turns this inequality into $\Phi(\rho)\le(1+t)\Phi(\sigma)$ with $\Phi(\sigma)\in\mathcal F$, so it cannot increase $R_{\mathcal F}$. This makes the robustness a resource measure for all four resources.
The generalized robustness also quantifies an advantage in the discrimination of quantum operations (subchannel discrimination). Over all such tasks, $1+R_{\mathcal F}(\rho)$ is the largest ratio between the success probability with input state $\rho$ and the best success probability with a free input state~\cite{Takagi2019}.

\paragraph{Coherence.}
Coherence is superposition with respect to a fixed basis, here the computational basis.
The free states are the diagonal states, also called incoherent, and free operations cannot create off-diagonal entries from them~\cite{Baumgratz2014,Streltsov2017}.
Coherence underlies interference and has been studied as a resource for quantum algorithms, metrology and thermodynamics~\cite{Streltsov2017}.
For a qubit, \eqref{eq:robustness} equals ${C(\rho)=\sum_{j\ne k}\abs*{\rho_{jk}}}$, the $\ell_1$-norm of coherence~\cite{Baumgratz2014,Napoli2016}.

\paragraph{Imaginarity.}
Imaginarity is defined with respect to the same basis.
The free states are the real states, whose density matrices have only real entries, $\rho=\rho^{\tp}$, and the free operations are those with real Kraus operators~\cite{HickeyGour2018,Wu2021Operational,Wu2021}.
The theory was introduced to study what complex numbers contribute to quantum information processing.
For example, the two-qubit states $(\id\pm Y\otimes Y)/4$ are real, and local operations with classical communication distinguish them perfectly. When the local measurements are restricted to real ones, the two states give identical outcome distributions~\cite[Eq.~(10)]{Wu2021Operational}.
Here \eqref{eq:robustness} is the robustness of imaginarity~\cite[Eq.~(61)]{Wu2021}, with the closed form ${I(\rho)=\frac12\norm*{\rho-\rho^{\tp}}_1}$~\cite[Proposition~5]{Wu2021}.

\paragraph{Magic.}
For magic the restriction is to stabilizer operations: Clifford gates, measurements in the computational basis and the preparation of stabilizer states~\cite{Veitch2014}. A Clifford gate maps Pauli operators to Pauli operators up to signs under conjugation, and the stabilizer states are the states that Clifford gates produce from computational basis states.
A computation built from these operations alone can be simulated efficiently on a classical computer~\cite{AaronsonGottesman2004}, and it becomes universal for quantum computation when suitable non-stabilizer states, called magic states, are supplied~\cite{BravyiKitaev2005}.
The pure stabilizer states of a qubit are the six eigenstates of the Pauli operators $X$, $Y$, $Z$, and the free states form their convex hull $\mathrm{STAB}$. We use the generalized robustness of magic $M=R_{\mathrm{STAB}}$, normalized to vanish on free states.
In Ref.~\cite[Eq.~(15)]{Rubboli2024} this measure appears as $\Lambda^+-1$, where $\Lambda^+(\rho)$ is the least $\lambda$ with $\rho\le\lambda\sigma$ for some $\sigma\in\mathrm{STAB}$.

\paragraph{Entanglement.}
A state of the two qubits is separable if it is a convex mixture of product states $\rho_A\otimes\rho_B$, and entangled otherwise~\cite{Horodecki2009}.
Two parties who each hold one of the qubits can prepare exactly the separable states by local operations and classical communication, and these operations cannot create entanglement.
Entangled states are the resource for teleportation, dense coding and quantum cryptography~\cite{Horodecki2009}.
We write $\mathrm{Sep}$ for the set of separable states and $R_g=R_{\mathrm{Sep}}$ for the generalized robustness of entanglement~\cite{Steiner2003}.
In the robustness of Vidal and Tarrach~\cite{VidalTarrach1999} the admixed state $\tau$ in~\eqref{eq:robustness} is separable; here it is arbitrary.

\subsection{Resource power}
\label{sec:power}
A gate acts on a state that depends on the circuit input and on the preceding gates. We quantify the gate independently of this state by its largest possible resource increase.
For any of the measures $R$ above, we use the \emph{resource increasing power} of a unitary $V$~\cite[Eq.~(2)]{LiBuLiu2020}, which we call its \emph{resource power} for short. It is the largest increase in $R$ over input states,
\begin{equation}
 p_R(V)=\max_\rho\bigl\{R(V\rho V^\dagger)-R(\rho)\bigr\},
\label{eq:resource-power}
\end{equation}
where the maximum is over all states of the system on which $V$ acts, without ancillas.
We write $p_C$, $p_I$, $p_M$ and $p_E$ for the cohering, imaginarity, magic and entangling power.
For coherence, \eqref{eq:resource-power} is the coherence power of Ref.~\cite[Definition~1]{GarciaDiaz2016}, called generalized cohering power in Ref.~\cite[Eq.~(6)]{Bu2017Cohering}.
Other definitions take the largest absolute change of the resource, $\max\{p_R(V),p_R(V^\dagger)\}$~\cite{Bu2024,Ye2026}, or the largest resource created from a free state. For all gates whose power we compute, they give the same value (Lemma~\ref{lem:powers} and Appendix~\ref{app:rzz-power}).
Averages over free input states are also used, for example the entangling power of Zanardi et al.~\cite{Zanardi2000} and the imaginarity-generating power of Ref.~\cite{Awasthi2026}.

The power is nonnegative, because $V$ leaves the free, maximally mixed state unchanged.
It vanishes exactly when $V$ maps free states to free states: a gate that takes some free state outside $\mathcal F$ has positive power, and a gate that preserves the free states cannot increase $R$ (Section~\ref{sec:measures}).

For a specified gate sequence, we define the \emph{cumulative resource power} as the sum of the individual gate powers,
\begin{equation}
 P_R=\sum_j p_R(V_j).
\label{eq:resource-budget}
\end{equation}
For any circuit input, each gate $V_j$ raises $R$ by at most $p_R(V_j)$, so the gates counted in~\eqref{eq:resource-budget} contribute a total increase of at most $P_R$.
In our models these are the trainable gates, not the encoding gates.
The cumulative power is a property of the gate sequence: different decompositions of the same unitary can have different cumulative powers (Appendix~\ref{app:power-dependence}).

\subsection{One qubit in the Bloch picture}
\label{sec:bloch}
\begin{figure}[t]
\centering
\includegraphics[width=\linewidth]{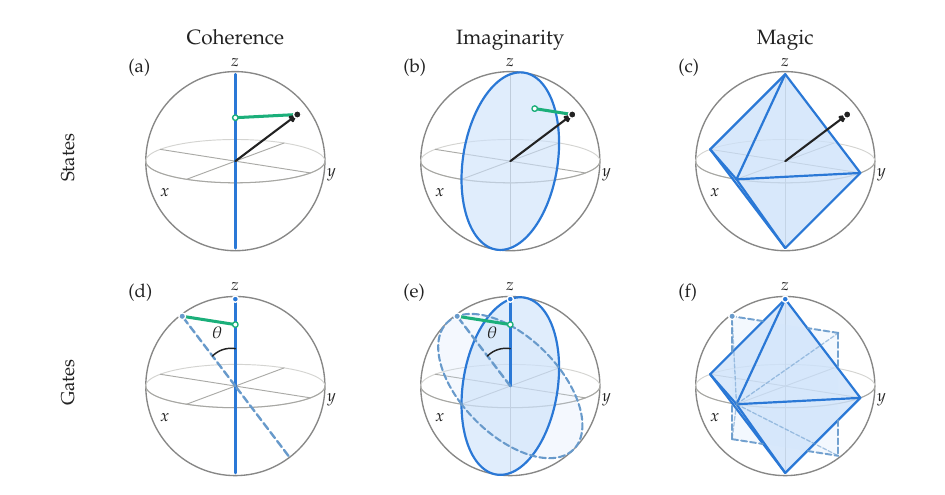}
\caption{One-qubit resources in the Bloch ball.
Panels (a)-(c) show the same state with Bloch vector \(s=(-1,1,1)/2\).
The free states form the blue \(z\) axis, \(xz\) disk and stabilizer octahedron, respectively.
In (a) and (b) the green segment joins the state to its orthogonal projection onto the free set, and its length is the coherence in (a) and the imaginarity in (b).
In (c) the state lies outside the stabilizer octahedron.
Panels (d)-(f) use the same gate \(R_x(\pi/4)\). Light blue dashes show the rotated free sets. The two blue dots mark the north pole before and after rotation.
In (d) and (e) the green segment has length \(\sin\theta\), where \(\theta=\pi/4\) is the marked tilt angle, giving the cohering and imaginarity powers, respectively.
In (f) the rotated north pole lies outside the original octahedron, so the gate has positive magic power.}
\label{fig:bloch}
\end{figure}

In the Bloch ball, one-qubit states are represented by vectors and unitary gates act as rotations. The free states of coherence, imaginarity and magic form an axis, a disk and an octahedron, respectively (Figure~\ref{fig:bloch}).

\paragraph{States.}
A state is $\rho=(\id+s\cdot\sigma)/2$ with $\sigma=(X,Y,Z)$ and a Bloch vector $s\in\mathbb R^3$, $\enorm{s}\le1$; pure states have $\enorm{s}=1$.
We write $e_x$, $e_y$, $e_z$ for the unit vectors along the axes and $s_\perp=(s_x,s_y)^{\tp}$ for the part of $s$ in the transverse ($xy$) plane.
The diagonal states lie on the $z$ axis and the real states in the $xz$ plane (Figure~\ref{fig:bloch}(a), (b)), and coherence and imaginarity are the distances of $s$ from these sets,
\begin{equation}
 C(\rho)=\enorm{s_\perp},\qquad I(\rho)=\abs*{s_y}.
\label{eq:bloch-measures}
\end{equation}
The pure stabilizer states are the points $\pm e_x$, $\pm e_y$, $\pm e_z$, and $\mathrm{STAB}$ is the octahedron $\abs*{s_x}+\abs*{s_y}+\abs*{s_z}\le1$ (Figure~\ref{fig:bloch}(c)).

\paragraph{Gates.}
We write $R_a(t)=e^{-\ii t\sigma_a/2}$ for rotations generated by the Pauli matrices. Up to a global phase, a one-qubit gate is a unitary matrix of determinant one, $V\in\mathrm{SU}(2)$.
Conjugation by $V$ acts on the Bloch vector by a real orthogonal matrix of determinant one, $O(V)\in\mathrm{SO}(3)$, the Bloch rotation of $V$,
\[
 V(s\cdot\sigma)V^\dagger=(O(V)s)\cdot\sigma.
\]
These rotations satisfy $O(VW)=O(V)O(W)$ and $O(V^\dagger)=O(V)^{\tp}$. The gate $R_a(t)$ acts as the rotation by the angle $t$ about $e_a$, as shown for $R_x(\pi/4)$ in Figure~\ref{fig:bloch}(d)-(f).
A Clifford gate maps $X$, $Y$, $Z$ to Pauli operators up to signs, so its Bloch rotation permutes the axes up to signs.

\paragraph{Powers.}
The cohering and imaginarity powers depend only on how far the Bloch rotation tilts the $z$ axis and the $y$ axis.
\begin{lemma}[Cohering and imaginarity power of qubit gates]
\label{lem:powers}
For $V\in\mathrm{SU}(2)$,
\begin{equation}
 \begin{aligned}
  p_C(V)&=\sqrt{1-O(V)_{zz}^2},\\
  p_I(V)&=\sqrt{1-O(V)_{yy}^2}.
 \end{aligned}
\label{eq:powers}
\end{equation}
For coherence, the largest increase occurs for the basis state $\ket{0}$, so $p_C(V)=C(V\dyad{0}V^\dagger)$. For imaginarity, it occurs for a pure real state with Bloch vector proportional to $(O(V)_{yx},0,O(V)_{yz})$, provided $p_I(V)>0$.
\end{lemma}
The proof is given in Appendix~\ref{app:qubit-powers}. For coherence, the lemma agrees with the results for the $\ell_1$-norm of coherence in Refs.~\cite[Proposition~9]{Bu2017Cohering} and~\cite[Theorem~1]{GarciaDiaz2016}.

Geometrically, $p_C(V)$ is the sine of the angle by which $O(V)$ tilts the $z$ axis (Figure~\ref{fig:bloch}(d)).
The transverse projections of $O(V)e_z$ and $O(V)^{\tp}e_z$ both have length $p_C(V)$, since both vectors have unit length and $z$ component $O(V)_{zz}$.
In terms of the matrix entries, $O(V)_{zz}=\abs*{V_{00}}^2-\abs*{V_{10}}^2$ is the $Z$ expectation value of $V\ket{0}$.
The two squares add up to one, so that $p_C(V)=2\abs*{V_{00}}\abs*{V_{10}}$, twice the product of the absolute values of a diagonal and an off-diagonal entry of $V$.
In the same way, $p_I(V)$ is the sine of the angle by which $O(V)$ tilts the $y$ axis, and with it the $xz$ plane of real states (Figure~\ref{fig:bloch}(e)).

This geometry also characterizes the gates of zero resource power.
A gate has zero cohering power exactly when its rotation maps the $z$ axis to itself, possibly reversing its direction.
It has zero imaginarity power exactly when the same holds for the $y$ axis, that is, when its matrix is real up to a global phase.
It has zero magic power exactly when it is a Clifford gate (Figure~\ref{fig:bloch}(f)).
Indeed, by Section~\ref{sec:power} such a gate maps the octahedron $\mathrm{STAB}$ into itself, and a rotation does so only if it maps the vertices, the only unit vectors in the octahedron, to vertices.

\subsection{Approximation error and nonaffinity}
\label{sec:approximation}
We compare a model function $f$ with a target function $g$ using the following notions of approximation.
For a compact input domain $\X$, $\cont(\X;J)$ denotes the continuous functions on $\X$ with values in $J$.
Unless stated otherwise, approximation is uniform. On a domain $D$ we write $\norm*{h}_{\infty,D}=\sup_{x\in D}\abs*{h(x)}$ and omit $D$ when the domain is clear.
The \emph{distance} of $g$ from a nonempty class $A$ of functions is the infimum of the error over the class,
\begin{equation}
 \dist(g,A)=\inf_{h\in A}\supnorm{g-h},
\label{eq:dist}
\end{equation}
and the uniform closure of $A\subseteq\cont(\X;\mathbb R)$ consists of all $g\in\cont(\X;\mathbb R)$ at distance zero from $A$.
The oscillation $\osc(h)=\max h-\min h$ is twice the distance of $h$ from the constant functions.
We write $(a)_+=\max\{a,0\}$ and $h_{\odd}(x)=[h(x)-h(-x)]/2$ for the odd part of $h$ on $[-1,1]$.

Section~\ref{sec:magic} also considers the mean absolute error $\norm*{f-g}_{L^1(\mu)}$, where the Borel probability measure $\mu$ describes the input distribution and
\begin{equation}
 \norm*{h}_{L^1(\mu)}=\int_{\X}|h(x)|\,d\mu(x).
\label{eq:mean-error}
\end{equation}

With phase encoding (Section~\ref{sec:model}) the model function is a trigonometric polynomial in $x$~\cite{Schuld2021}, that is, a finite sum of terms $c_ke^{\ii kx}$ with integers $k$; its degree is the largest $\abs*{k}$ that occurs. It therefore has an absolutely convergent Chebyshev expansion on $[-1,1]$~\cite[Chap.~3]{Trefethen2019}. We write
\begin{equation}
 \begin{aligned}
  f(x)&=\sum_{n\ge0}a_nT_n(x),\\
  \N(f)&=\sum_{n\ge2}\abs*{a_n},
 \end{aligned}
\label{eq:cheb}
\end{equation}
where $T_n(\cos\varphi)=\cos(n\varphi)$ and $a_0$ carries no factor $1/2$.
Since $T_0(x)=1$ and $T_1(x)=x$, $\N(f)$ vanishes exactly when $f$ is affine. We therefore use it as a measure of \emph{nonaffinity}.
The bound $\abs*{T_n}\le1$ on $[-1,1]$ implies that the distance from $f$ to the affine functions is at most $\N(f)$, but the two quantities need not coincide.

\section{The one-qubit model}
\label{sec:model}
Universal approximation with a single qubit was established by P\'erez-Salinas et al.~\cite[Theorem~4]{PerezSalinas2021} for trainable input weights and by Yu et al.~\cite{Yu2022} for a fixed encoding rotation. We work in the second setting.

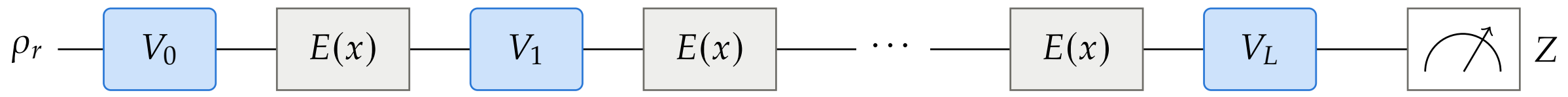
\begin{figure}[ht]
\centering
\resizebox{\linewidth}{!}{\begin{tikzpicture}[
  line width=0.5pt,
  control/.style={draw=figblue, fill=figbluelight, rounded corners=2pt,
                  minimum width=11mm, minimum height=8mm, inner sep=2pt},
  encoder/.style={draw=figmuted, fill=figshade!45,
                  minimum width=13mm, minimum height=8mm, inner sep=2pt},
  meter/.style={draw=figmuted, fill=white,
                minimum width=11mm, minimum height=8mm, inner sep=0pt}
]
  \definecolor{figblue}{HTML}{2A78D6}
  \definecolor{figbluelight}{HTML}{CDE2FB}
  \definecolor{figmuted}{HTML}{6F6E69}
  \definecolor{figshade}{HTML}{D9D9D4}

  \draw (0.6,0) -- (14.4,0);

  \node[left] at (0.6,0) {$\rho_r$};

  \node[control] at (1.6,0)  {$V_0$};
  \node[encoder] at (3.4,0)  {$E(x)$};
  \node[control] at (5.2,0)  {$V_1$};
  \node[encoder] at (7.0,0)  {$E(x)$};
  \node[fill=white, inner sep=4pt] at (8.8,0) {$\cdots$};
  \node[encoder] at (10.6,0) {$E(x)$};
  \node[control] at (12.4,0) {$V_L$};

  \node[meter] (m) at (14.4,0) {};
  \draw ([xshift=1.8mm,yshift=1.9mm]m.south west) arc[start angle=180, end angle=0, radius=3.7mm];
  \draw[->] ([yshift=1.9mm]m.south) -- ++(2.6mm,4.4mm);
  \node[right] at (m.east) {$Z$};
\end{tikzpicture}}
\caption{Data re-uploading circuit~\eqref{eq:circuit} with data-independent trainable gates $V_j$, a fixed encoding gate $E(x)$ and $Z$ readout.}
\label{fig:reuploading}
\end{figure}

Let $x\in\X$ be the input. The encoding gate $E(x)$ is applied $L\in\mathbb N_0$ times. We call each application an encoding call and $L$ the encoding depth, or simply the depth. With the parameters $\params=(\theta_0,\ldots,\theta_L)$ of the $L+1$ trainable gates, the circuit and its output are
\begin{align}
 \rho_r&=\frac12(\id+rZ),\qquad 0<r\le1,\notag\\
 U_{\params}(x)&=V_LE(x)\cdots V_1E(x)V_0,
\label{eq:circuit}\\
 \rho_{\params}(x)&=U_{\params}(x)\rho_rU_{\params}(x)^\dagger,\notag\\
 f_{\params}(x)&=\Tr{Z\rho_{\params}(x)},
\label{eq:model}
\end{align}
with $U_{\params}(x)=V_0$ for $L=0$ (Figure~\ref{fig:reuploading}). The measured observable, here $Z$, is the readout, and its expectation value $f_{\params}$ is the model function.
The data-independent trainable gates $V_j=V(\theta_j)\in\mathrm{SU}(2)$ are restricted only through their cumulative resource power.
For a fixed circuit we drop the index $\params$ and write $U$, $\rho$ and $f$.
The input state $\rho_r=(1-r)\id/2+r\dyad{0}$ is the pure state $\ket{0}$ mixed with the maximally mixed state, and its Bloch vector has length $r$. Values $r<1$ model depolarizing noise on the input. Since $Z$ is traceless, the maximally mixed component contributes zero to $f$. What remains is $r$ times the model function for the pure input $\ket{0}$, so $f$ takes values in $[-r,r]$.

Table~\ref{tab:encodings} gives the three encoding gates and their domains. Sections~\ref{sec:local} and~\ref{sec:magic} use phase encoding.
\begin{table}[ht]
\caption{Fixed encoding gates and their input domains. The amplitude encoding gate maps $\ket{0}$ to $\sqrt{1-x}\ket{0}+\sqrt{x}\ket{1}$, and each encoding call applies this entire gate.}
\label{tab:encodings}
\begin{center}
\begin{tabular}{lll}
\toprule
Encoding & Gate $E(x)$ & Input domain $\X$\\
\midrule
$z$-rotation (phase) & $R_z(x)$ & $[-1,1]$\\
$y$-rotation & $R_y(x)$ & $[-1,1]$\\
Amplitude & $R_y(2\arcsin\sqrt{x})$ & $[0,1]$\\
\bottomrule
\end{tabular}
\end{center}
\end{table}

With unrestricted trainable gates and depth, the model with any of these encodings approximates every target in $\cont(\X;[-r,r])$ uniformly (Proposition~\ref{prop:one-qubit-universality}).
This known universality provides the reference for our bounds: without resource restrictions the model can approximate every target, so the bounds identify limitations imposed by those restrictions.
Yu et al.\ characterize the unitaries that the circuits implement with phase encoding by Laurent polynomials, that is, polynomials in $e^{\ii x/2}$ and $e^{-\ii x/2}$, for general and for real trainable gates~\cite[Lemmas~3 and~1]{Yu2022}, and deduce approximation theorems. In the published version these are formulated for square-integrable targets on the full period $[-\pi,\pi]$ and up to a normalizing constant~\cite[Theorem~4 and Proposition~2]{Yu2022}.
The proofs of Theorems~\ref{thm:i1}--\ref{thm:magic-mean} require uniform approximation of the target itself on the stated input intervals, with real trainable gates for even targets.
Appendix~\ref{app:synthesis} derives these statements from the two lemmas of Yu et al.\ using Fej\'er's theorem and the Fej\'er--Riesz factorization, the polynomial approximation results collected in Appendix~\ref{app:approximation-tools}.

The model function also determines the distribution of the measurement outcome, so approximating a function amounts to approximating a conditional distribution.
At fixed $x$, measuring $Z$ gives bit $y=0$ for eigenvalue $+1$ and $y=1$ for $-1$. The conditional probabilities are $p(0\mid x)=[1+f(x)]/2$ and $p(1\mid x)=[1-f(x)]/2$~\cite[Eq.~(14)]{PerezSalinas2021}.
For a target conditional distribution $p_*(\cdot\mid x)$ on $\{0,1\}$, let $g(x)=2p_*(0\mid x)-1$. The two probability errors have opposite signs and equal magnitude. The total variation distance between the distributions, $\operatorname{TV}(p,q)=\tfrac12\sum_y|p(y)-q(y)|$, therefore equals $\tfrac12|f(x)-g(x)|$ at each $x$. Taking the supremum or averaging over the input distribution $\mu$ gives
\begin{align}
 \sup_{x\in\X}\operatorname{TV}\!\left(p(\cdot\mid x),p_*(\cdot\mid x)\right)
 &=\tfrac12\supnorm{f-g},
\label{eq:generative}\\
 \int_{\X}\operatorname{TV}\!\left(p(\cdot\mid x),p_*(\cdot\mid x)\right)\,d\mu(x)
 &=\tfrac12\norm*{f-g}_{L^1(\mu)}.
\label{eq:generative-mean}
\end{align}
With unrestricted trainable gates and depth, the uniformly approximable conditional distributions on $\{0,1\}$ are precisely those for which $x\mapsto p_*(0\mid x)$ is continuous and takes values in $[(1-r)/2,(1+r)/2]$.

\section{Coherence and imaginarity in the one-qubit model}
\label{sec:local}
We use phase encoding $R_z(x)$ on $[-1,1]$ and measure $Z$. The cumulative powers $P_C$ and $P_I$ sum the cohering and imaginarity powers~\eqref{eq:resource-power} of the trainable gates $V_0,\ldots,V_L$ in~\eqref{eq:circuit}.

The initial Bloch vector is $re_z$, and the readout is its final $z$ component. In the notation of Section~\ref{sec:bloch},
\begin{equation}
 \begin{aligned}
  f(x)&=r\,e_z^{\tp}\,O(V_L)R(x)\cdots O(V_1)R(x)\,O(V_0)\,e_z,\\
  R(x)&=\begin{pmatrix}Q_x&0\\0&1\end{pmatrix},
 \end{aligned}
\label{eq:bloch-response}
\end{equation}
where $R(x)=O(R_z(x))$ is the rotation by the angle $x$ about the $z$ axis and $Q_x$ rotates the transverse plane by this angle.

\subsection{Coherence and oscillation}
\label{sec:coherence}
The phase gate rotates the transverse component of the Bloch vector and leaves its $z$ component unchanged. For $f$ to depend on $x$, a trainable gate must first create a transverse component and a later gate must rotate part of it into the $z$ direction. A gate $V$ maps $e_z$ to a unit Bloch vector whose transverse component has length $p_C(V)$. In the reverse direction, $p_C(V)$ is the largest absolute $z$ component obtained by rotating a unit vector in the transverse plane (Lemma~\ref{lem:powers} and Figure~\ref{fig:bloch}(d)). Consider one encoding call between two opposite rotations about $y$,
\begin{equation}
 \begin{aligned}
  U(x)&=R_y(-\alpha)R_z(x)R_y(\alpha),\\
  f(x)&=r\bigl(\cos^2\alpha+\sin^2\alpha\cos x\bigr).
 \end{aligned}
\label{eq:coherence-example}
\end{equation}
The first rotation sends $re_z$ to $r(\sin\alpha,0,\cos\alpha)$. The phase gate turns its transverse component through $x$, after which $R_y(-\alpha)$ converts part of it back into the measured component. Each trainable gate has cohering power $\abs*{\sin\alpha}$ by Lemma~\ref{lem:powers}, and the coefficient of $\cos x$ in $f/r$ is the product of these powers.
The general bound is quadratic in $P_C$ for the same reason.

\begin{theorem}[Cohering power bounds oscillation and nonaffinity]
\label{thm:c1}
Let $f$ be the model function of a circuit with $L\ge1$ encoding calls and cumulative cohering power $P_C$. Then
\begin{equation}
 \begin{aligned}
  \osc(f)&\le 2r\min\{1,P_C^2/4\},\\[2pt]
  \N(f)&\le K_L\,r\min\{1,P_C^2/4\},
 \end{aligned}
\label{eq:coherence}
\end{equation}
where $K_L=\min\{L^2/2,\sqrt{2\pi L}\}$.
When $P_C=0$ the only model functions are the constants $f\equiv r$ and $f\equiv-r$.
The bound on the oscillation is attained for every $P_C\le2$ when $L\ge2$.
\end{theorem}
The proof in Appendix~\ref{app:coherence} follows the two amplitudes of the qubit through the circuit with pure input $\ket{0}$.
Let $S$ be the largest product of the absolute values of the $\ket{0}$ amplitude at an input $x$ and the $\ket{1}$ amplitude at an input $y$, over all pairs of inputs.
The encoding gate changes only phases and leaves $S$ unchanged, whereas a trainable gate $V_j$ can increase it by at most $p_C(V_j)/2$.
Since $S$ starts at zero and the outputs at two inputs differ by at most $2rS^2$, the oscillation is at most $rP_C^2/2$.
The circuit~\eqref{eq:coherence-example}, with two or more encoding calls and an additional fixed phase $R_z(\pi/2)$, attains this bound.
To obtain the nonaffinity bound, we use that $f$ is a trigonometric polynomial of degree at most $L$. The proof bounds its oscillation over a full period; Bernstein's inequality then controls its derivatives and hence the sum of the absolute values of its Chebyshev coefficients beyond degree one.

The oscillation bound gives a necessary cohering power for approximating a target.
If a circuit approximates a target $g$ with $\supnorm{f-g}\le\varepsilon$, then $\osc(g)\le\osc(f)+2\varepsilon\le rP_C^2/2+2\varepsilon$, that is,
\begin{equation}
 P_C\ge\sqrt{\frac2r\bigl(\osc(g)-2\varepsilon\bigr)_+}.
\label{eq:coherence-target}
\end{equation}
At any depth, a nonconstant target requires positive $P_C$ once $\varepsilon<\osc(g)/2$. For a target with range $[-r,r]$, the required power is at least $2\sqrt{1-\varepsilon/r}$ when $0\le\varepsilon\le r$.
A single gate has cohering power at most one, so as $\varepsilon\to0$ this requirement approaches the combined power of two maximally cohering gates.
Exact realization of $r\cos4x$, which for $r=1$ is the example of the introduction, requires $P_C\ge2$.
This value is attained: in the circuit $R_y(-\pi/2)R_z(x)^4R_y(\pi/2)$ each of the two rotations about the $y$ axis has cohering power one, and~\eqref{eq:coherence-example} gives $f(x)=r\cos4x$.

\begin{figure}[t]
\centering
\includegraphics[width=\linewidth]{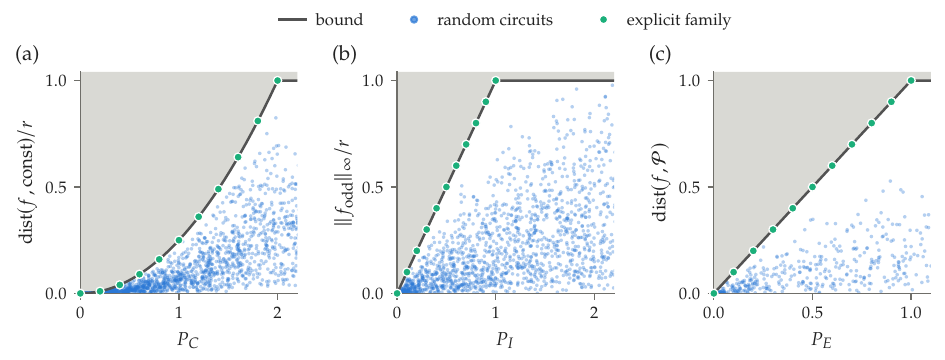}
\caption{Resource bounds and sampled circuits.
(a) Distance of $f$ from the constant functions, with the bound $\min\{1,P_C^2/4\}$ of Theorem~\ref{thm:c1}.
(b) Uniform norm of the odd part of $f$, with the bound $\min\{1,P_I\}$ of Theorem~\ref{thm:i1}.
(c) Distance of $f$ from the product functions, with the bound $\min\{1,P_E\}$ of Theorem~\ref{thm:entangling-power}. For the random circuits it is a numerical estimate: On a grid of $33\times33$ equally spaced inputs, we minimize the maximum error over the two factors of a product function, one at a time, until it improves by less than $10^{-7}$. We keep the smallest error from eight starting values.
The gray regions are excluded by the bounds. The explicit families saturate them (Appendices~\ref{app:coherence}, \ref{app:imaginarity} and~\ref{app:product-sharpness}).
The random circuits have up to eight encoding calls in (a) and (b) and up to three $R_{ZZ}$ gates in (c).}
\label{fig:bounds}
\end{figure}

The distance of $f$ from the constants is $\osc(f)/2$, the quantity shown in Figure~\ref{fig:bounds}(a). Since constants are affine, the distance from $f$ to the affine functions is also at most $rP_C^2/4$ at any depth. The nonaffinity bound controls the sum of the absolute values of all Chebyshev coefficients beyond degree one, but its constant $K_L$ grows with depth.

\subsection{Imaginarity and the odd part}
\label{sec:imag}
Complex conjugation in the computational basis reflects the Bloch vector in the $xz$ plane, the plane of real states (Figure~\ref{fig:bloch}(b)).
It leaves the input state $\rho_r$ and the readout $Z$ unchanged and maps $R_z(x)$ to $R_z(-x)$.
Real trainable gates therefore give an even model function. With unrestricted depth, such circuits uniformly approximate every continuous even target $g:[-1,1]\to[-r,r]$ (Proposition~\ref{prop:one-qubit-universality}).
For a general gate $V$, the Bloch rotations of $V$ and of its complex conjugate $\overline V$ differ in operator norm by exactly $2p_I(V)$ (Appendix~\ref{app:imaginarity}).
Summed over the trainable gates, these differences bound the odd part
\[
 f_{\odd}(x)=\frac{f(x)-f(-x)}2.
\]

\begin{theorem}[Imaginarity power bounds the odd part]
\label{thm:i1}
For phase encoding and $Z$ readout,
\begin{equation}
 \supnorm{f_{\odd}}\le r\min\{1,P_I\}.
\label{eq:imagbound}
\end{equation}
Allowing arbitrary encoding depth, the uniform closure of the model functions with $P_I=0$ consists of all even functions in $\cont([-1,1];[-r,r])$. For every $g\in\cont([-1,1];[-r,r])$, the infimum of $\supnorm{f-g}$ over these circuits equals $\supnorm{g_{\odd}}$.
The bound~\eqref{eq:imagbound} is attained for every $P_I\le1$ when $L\ge2$.
\end{theorem}
Since $\supnorm{g_{\odd}}\le\supnorm{f_{\odd}}+\supnorm{g-f}$, an approximation with $\supnorm{f-g}\le\varepsilon$ requires
\begin{equation}
 P_I\ge\frac1r\bigl(\supnorm{g_{\odd}}-\varepsilon\bigr)_+.
\label{eq:imag-target}
\end{equation}
The odd part includes any linear term in $g$. For an odd target the bound reduces to $P_I\ge(\supnorm{g}-\varepsilon)_+/r$. Appendix~\ref{app:imaginarity} gives circuits attaining~\eqref{eq:imagbound}, plotted in Figure~\ref{fig:bounds}(b).

\section{Two-qubit entanglement and product functions}
\label{sec:multi}
With two inputs $x_1$ and $x_2$, we ask how far the model function is from a product of functions of the individual inputs.
The two qubits start in $\ket{00}$. Qubit $i$ encodes $x_i\in D_i$ with one of the gates in Table~\ref{tab:encodings}, where $D_i$ is the input domain of this gate, and it has no direct access to the other input. A circuit is any finite sequence of these encoding gates, arbitrary local gates $U_A\otimes U_B$, and two-qubit gates $R_{ZZ}(\gamma)=e^{-\ii\gamma Z\otimes Z/2}$. Only the encoding gates depend on the inputs.

The readout is $Z\otimes Z$, which measures the parity of the two outcome bits. Its expectation value $f(x_1,x_2)$ is the model function, and $[1+f(x_1,x_2)]/2$ is the probability of even parity.

The gates $R_{ZZ}$ are the only gates of the model that can create entanglement, and their entangling power has a closed form.

\begin{lemma}[Entangling power of $R_{ZZ}$]
\label{lem:rzz-power}
For every $\gamma\in\mathbb R$, the entangling power~\eqref{eq:resource-power} of $R_{ZZ}(\gamma)$ is
\begin{equation}
 p_E(R_{ZZ}(\gamma))=\abs*{\sin\gamma},
\label{eq:native-entangling-power}
\end{equation}
and the maximum increase in $R_g$ is attained on a product state.
\end{lemma}
Appendix~\ref{app:rzz-power} proves the lemma. Local gates have zero entangling power, so the cumulative entangling power of a circuit with the two-qubit gates $R_{ZZ}(\gamma_1),\ldots,R_{ZZ}(\gamma_m)$ is
\[
 P_E=\sum_{j=1}^m\abs*{\sin\gamma_j}.
\]

With unrestricted depth and entangling power, the model approximates every $g\in\cont(D_1\times D_2;[-1,1])$ uniformly (Proposition~\ref{prop:two-qubit-universality}).
The model can therefore approximate targets that are not products; the bound below quantifies the entangling power they require.
To our knowledge, universality has been established for related models but not for this one. Appendix~\ref{app:two-qubit-universality} proves the proposition and compares these settings~\cite{PerezSalinas2021,Schuld2021,Manzano2025,Yu2024,PerezSalinas2024Minimal}. In the proof, three $R_{ZZ}(\pi/2)$ gates and local gates implement a SWAP, giving the first qubit access to both inputs. The construction then approximates a rotation whose angle is a trigonometric polynomial in the two encoding angles. The proof of Theorem~\ref{thm:entangling-power} does not depend on this universality result.

If $P_E=0$, every $R_{ZZ}$ angle is a multiple of $\pi$. Up to a global phase, $R_{ZZ}(k\pi)$ is the identity for even $k$ and $Z\otimes Z$ for odd $k$. The full circuit therefore factors into two local circuits, and its $Z\otimes Z$ expectation is the product of their $Z$ expectations.
This is why we compare $f$ with the class of bounded product functions,
\begin{equation}
 \mathcal P=\{(x_1,x_2)\mapsto a(x_1)b(x_2):
 a\in\cont(D_1;[-1,1]),\ b\in\cont(D_2;[-1,1])\}.
\label{eq:entangling-budget}
\end{equation}
\begin{theorem}[Entangling power bounds the distance from product functions]
\label{thm:entangling-power}
Every function $f$ computed by this model satisfies
\begin{equation}
 \dist(f,\mathcal P)\le\min\{1,P_E\}.
\label{eq:entangling-bound}
\end{equation}
Allowing arbitrarily many local encoding calls, the uniform closure of the model functions with $P_E=0$ is exactly $\mathcal P$.
For any number of applications of the local encoding gates, an approximation of $g\in\cont(D_1\times D_2;[-1,1])$ with $\supnorm{f-g}\le\varepsilon$ requires
\begin{equation}
 P_E\ge\bigl(\dist(g,\mathcal P)-\varepsilon\bigr)_+.
\label{eq:entangling-target}
\end{equation}
The bound~\eqref{eq:entangling-bound} is attained for every $P_E\le1$.
\end{theorem}
The proof in Appendix~\ref{app:product-distance} removes the $R_{ZZ}$ gates one at a time. Local gates first reduce each angle to $0\le\gamma_j\le\pi/2$. In this range, removing $R_{ZZ}(\gamma_j)$ and multiplying the resulting model function by $\cos\gamma_j$ changes the output by at most $\sin\gamma_j$. After all $R_{ZZ}$ gates have been removed, the circuit contains only local gates and its output is a product function. The accumulated error is at most $P_E$.
The factors $a$ and $b$ are arbitrary, so each may be highly nonaffine in its own input; the bound concerns only the lack of product structure.

For example, the target $\cos2x_1\cos2x_2+p\sin2x_1\sin2x_2$ on $[-1,1]^2$ has distance $p$ from $\mathcal P$ for $0\le p\le1$ (Appendix~\ref{app:product-sharpness}). Approximating it with a uniform error of at most $\varepsilon$ requires $P_E\ge p-\varepsilon$. At $p=1$ the target is $\cos(2x_1-2x_2)$. Figure~\ref{fig:bounds}(c) shows circuits attaining the distance bound for each $p$. The full class of functions approximable at a given positive $P_E$ remains unknown.

\section{Approximation with Clifford trainable gates}
\label{sec:magic}
We return to the one-qubit model of Section~\ref{sec:model}, with $R_z(x)$ as the encoding gate on $[-1,1]$ and $Z$ as the readout.
The bounds on oscillation, the odd part and distance from product functions hold at any depth. For magic the situation is different.
Restricting the trainable gates to Clifford gates is equivalent to requiring zero cumulative magic power $P_M$ (Section~\ref{sec:bloch} and~\eqref{eq:resource-power}), but on this interval the encoding gate itself is non-Clifford for every $x\ne0$.
Non-Clifford operations are therefore available even at $P_M=0$. We ask how much additional depth is needed when all trainable gates are Clifford.

\begin{theorem}[Clifford trainable gates suffice]
\label{thm:magic-mean}
Let $g\in\cont([-1,1];[-r,r])$ and $\varepsilon,\delta\in(0,1)$.
There is a circuit~\eqref{eq:circuit} with phase encoding and Clifford trainable gates, so that $P_M=0$, whose model function satisfies
\begin{equation}
 \abs*{f(x)-g(x)}\le\varepsilon
 \qquad\text{whenever }\delta\le\abs*{x}\le1.
\label{eq:magic-mean}
\end{equation}
If $g$ is the model function of a circuit with $D$ encoding calls and unrestricted trainable gates, the number of encoding calls can be chosen polynomial in $D$, $1/\varepsilon$ and $1/\delta$.
\end{theorem}
The proof in Appendix~\ref{app:magic} starts from a circuit with unrestricted trainable gates. It approximates each of these gates by a sequence of Clifford gates and encoding gates.
The construction uses a rotation by $\pi/2$ about an axis at angle $\pi/4$ to the $z$ axis. This gate is not Clifford and, together with the Clifford gates, generates a dense set of one-qubit gates.
A recursion builds such a rotation from Clifford gates and encoding gates $R_z(x)$: the angle between its axis and the $z$ axis converges geometrically to $\pi/4$, uniformly for $\delta\le\abs*{x}\le1$.
The direction of the axis around the $z$ axis, its azimuth $\phi(x)$, depends on the input and is not controlled, but it is the same for all gates built from this rotation.
Each trainable gate $V_j$ is then approximated only up to a common rotation, by a gate sequence $W_j(x)$ close to $B(x)V_jB(x)^\dagger$ with $B(x)=R_z(\phi(x))$.
Since $B(x)$ commutes with the encoding gate, the rotations between successive gates cancel. The two that remain at the ends commute with the input state and the readout, so the model function does not change.

\begin{figure}[t]
\centering
\includegraphics[width=\linewidth]{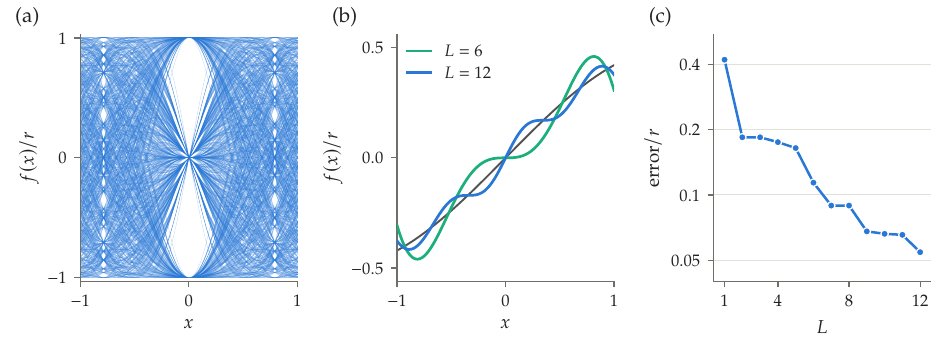}
\caption{Approximation with Clifford trainable gates.
(a) All distinct model functions with at most five encoding calls, 863 in total.
(b) The target $g(x)=(r/2)\sin x$ (gray) and its best uniform approximations with at most $L=6$ and $L=12$ encoding calls.
(c) Smallest uniform error on $[-1,1]$ for this target with at most $L$ encoding calls, normalized by $r$.
The circuits are enumerated exhaustively, and their errors are evaluated numerically on a fine grid.}
\label{fig:magic-approximation}
\end{figure}

Theorem~\ref{thm:magic-mean} excludes a neighborhood of $x=0$ because the encoding gate is the identity there: the whole circuit is then a Clifford gate, and $f(0)\in\{-r,0,r\}$.
This obstruction does not prevent approximation in mean when the input distribution $\mu$ gives probability zero to the single input $x=0$, that is, $\mu(\{0\})=0$. The probability of the excluded interval then tends to zero as $\delta\to0$. Since $\abs*{f-g}\le2r$ there, the theorem gives
\[
 \norm*{f-g}_{L^1(\mu)}\le\varepsilon+2r\,\mu\bigl(\{x:\abs*{x}<\delta\}\bigr).
\]
Both terms on the right can be made arbitrarily small, and with them the mean error.
By~\eqref{eq:generative-mean} the same conclusion holds for the mean total variation error when $p_*(0\mid x)$ is continuous in $x$ and takes values in $[(1-r)/2,(1+r)/2]$. For $r=1$ this is the entire probability interval.

At fixed depth, non-Clifford trainable gates can nevertheless improve the approximation, as an example with one encoding call shows.
The first Clifford gate sends $re_z$ to one of the six vectors $\pm re_x,\pm re_y,\pm re_z$. The encoding $R_z(x)$ rotates the transverse components through $x$ and preserves the $z$ component. The final Clifford gate selects one of the three resulting components, up to sign, as the readout (Section~\ref{sec:bloch}). The possible model functions are therefore $0$, $\pm r$, $\pm r\cos x$ and $\pm r\sin x$.
For example, $R_y(\pi/2)$ followed by $R_z(x)$ gives the Bloch vector $r(\cos x,\sin x,0)$. Under a final rotation $R_x(\theta)$, the incoming $y$ component contributes to the measured $z$ component with coefficient $\sin\theta$, so $f(x)=r\sin\theta\sin x$.
For $\theta=\pi/6$ this is $(r/2)\sin x$. Every model function with Clifford trainable gates and one encoding call differs from this target by at least $(r/2)\sin1$ in the uniform norm.
A non-Clifford trainable gate, by contrast, realizes this target exactly with one encoding call. With Clifford trainable gates, the smallest uniform error drops from $(r/2)\sin1\approx0.42\,r$ at one encoding call to $0.054\,r$ at twelve (Figure~\ref{fig:magic-approximation}).

\section{Discussion and outlook}

\label{sec:programme}
For data re-uploading models with fixed encoding and readout, we have related the resource powers of the trainable gates to properties of the model function.
With phase encoding on one qubit, cohering power limits its oscillation and imaginarity power its odd part.
For two qubits that each encode one input, entangling power limits its distance from product functions.
The three bounds are sharp and hold at any depth, so additional layers cannot make up for a missing resource.
They give lower bounds on the cumulative power required to approximate a target (Table~\ref{tab:resource-summary}).
Only the bound on the nonaffinity depends on the depth, and its constant $K_L$ is not known to be optimal.
For magic, the phase encoding itself supplies non-Clifford operations. On domains bounded away from $x=0$, Clifford trainable gates suffice for uniform approximation, with the polynomial increase in depth stated in Theorem~\ref{thm:magic-mean}.
Magic in the trainable gates can improve the accuracy at a fixed depth, but on such domains it is not required.

The requirements depend on the encoding and readout as well as on the target function.
With phase encoding, both $R_z(x)$ and $Z$ are diagonal in the reference basis, so trainable gates must convert between diagonal and off-diagonal state components before the input can affect the output.
With the encoding $R_y(x)$ and identity trainable gates, however, $U(x)=R_y(Lx)$ gives $f(x)=r\cos(Lx)$ at $P_C=0$, as in the example of the introduction: the encoding gate itself supplies the coherence.
The requirements also depend on the elementary gate set.
If SWAP were an elementary gate, it would have zero entangling power but could move the state between the two input wires and produce $\cos(x_1+x_2)$, which is not a product.
Our model instead decomposes SWAP into local gates and $R_{ZZ}$ gates whose powers enter $P_E$.
Splitting one $R_{ZZ}$ rotation into smaller rotations can likewise change this sum (Appendix~\ref{app:power-dependence}).
Each bound is therefore a requirement on the trainable gates of a specified model, with its encoding, readout and elementary gates.

The bounds also apply to generative learning when the model produces samples as measurement outcomes with conditional probabilities $p(y\mid x)$.
For one qubit, the error of the model function is twice the total variation error of the conditional Bernoulli distribution, by Eqs.~\eqref{eq:generative} and~\eqref{eq:generative-mean}.
The resource lower bounds therefore apply directly to these conditional distributions.
For several qubits, measurement produces bit strings~\cite{LiuWang2018}, and a single moment such as $\expval{Z\otimes Z}$ no longer determines their joint distribution.
Other generative models treat a vector of expectation values, evaluated at a random input, as a real-valued sample~\cite{Barthe2025Generative}.
Approximating a particular map is then sufficient to approximate its output law, because the Wasserstein-1 distance $W_1$ between two such laws is at most the mean distance between the two maps at the same random input~\cite[Chap.~6]{Villani2009}.
A resource lower bound for that map, however, need not apply to every map generating the same law.
In the product encoding of Barthe et al., for example, each output component comes from a separate one-qubit circuit with access to the full common input; this shared randomness can correlate the outputs without entanglement.

\paragraph{Open questions.}
\begin{enumerate}[leftmargin=*,itemsep=4pt,topsep=3pt,beginpenalty=10000]
 \item \textbf{Functions and resources.}
 Other encodings, measurements and multiqubit resources, including contextuality~\cite{Bowles2023}, require their own function-class characterizations. Block encoding and QSVT~\cite{Gilyen2019}, alongside other protocols for nonlinear transformations~\cite{Holmes2023}, provide further settings in which to study resource requirements and the functions approximable with free operations.
 For magic, our construction with Clifford trainable gates has polynomial depth, but it is not optimized. The optimal depth, and the gain from non-Clifford trainable gates at a fixed depth, are open.
 A broader question is which function classes admit simple representations by quantum models.
 \item \textbf{Training.}
We seek resource-dependent lower bounds on the number of statistical queries needed to learn with quantum models.
We conjecture that, among suitable ensembles of target functions, those whose targets have greater resource requirements have higher average-case statistical query complexity at fixed learning accuracy and query tolerance.
Here, the resource requirement of a target is the minimum cumulative resource power needed to approximate it to this accuracy within a fixed family of quantum learning models.
This conjecture is motivated by average-case hardness results in the statistical query model for learning the output distributions of random brickwork circuits~\cite{Nietner2025} and of random circuits from the orthogonal and symplectic groups~\cite{Shaya2026}. Haar random states are known to be rich in several resources~\cite{Deneris_2026}, including entanglement~\cite{Hayden2006}, coherence~\cite{SinghZhangPati2016}, magic~\cite{TurkeshiTirritoSierant2025}, imaginarity~\cite{Haug2025} and, on more than half of the system, fermionic non-Gaussianity~\cite{AresMurcianoCalabrese2026}.
These results motivate the conjecture but do not establish it. They concern the resources contained in random states rather than the power needed to approximate a target, and they use different resource measures from ours. Hardness also does not require every resource: random orthogonal circuits are hard to learn although their states are real and thus free of imaginarity.

 \item \textbf{Generalization.}
 Resource-dependent statistical complexity bounds~\cite{Bu2023Statistical} provide a starting point for studying generalization. We ask when additional resources improve performance on unseen data and what resource levels are needed at a given sample size. The error due to finite training data must be distinguished from approximation error.
\end{enumerate}

\section*{Acknowledgments}
The authors thank Eric R. Anschuetz for discussions.
During this
work, O.S. was supported by the Federal Ministry of Research, Technology
and Space (BMFTR), Germany, through the QC service center QUICS (grant
no.~13N17418). L.P. is supported by the National Research Foundation, Singapore through the National Quantum Office, hosted in A*STAR, under its Centre for Quantum Technologies Funding Initiative (S24Q2d0009). L.P. is also partially supported by the A*STAR under its Young Investigator Research Grant (YIRG) M25N8c0131.

\section*{Declaration of AI use}
For literature review, derivations and manuscript preparation, the authors first used OpenAI's GPT-6 Astra and then Anthropic's Claude Fable~5, reviewing both models' contributions line by line.

\appendix
\numberwithin{equation}{section}

\phantomsection
\addcontentsline{toc}{section}{Appendix}
\section*{Appendix}
The appendix contains the proofs. Appendix~\ref{app:preliminaries} provides auxiliary results, Appendix~\ref{app:synthesis} proves the universality of the one-qubit model, and Appendices~\ref{app:coherence}--\ref{app:magic} prove Theorems~\ref{thm:c1}--\ref{thm:magic-mean} in order.

\section{Auxiliary results}
\label{app:preliminaries}
This appendix collects the circuit approximation bounds, the proof of Lemma~\ref{lem:powers}, and the results on trigonometric and Chebyshev polynomials used below.

\subsection{Circuit approximation bounds}
\label{app:matrix-estimates}
We use two standard estimates that relate errors in individual gates to errors in the implemented unitary and in expectation values~\cite[Sec.~4.5.3, Box~4.1]{NielsenChuang2010}. We state the product estimate for matrices of operator norm at most one, and the measurement estimate for arbitrary input states and observables.

Let the square matrices $A_j,B_j$ have the same size and operator norm at most one. We use the telescoping identity
\[
 A_m\cdots A_1-B_m\cdots B_1
 =\sum_{j=1}^m A_m\cdots A_{j+1}(A_j-B_j)B_{j-1}\cdots B_1,
\]
where an empty product is the identity. The triangle inequality and submultiplicativity of the operator norm give
\begin{equation}
 \opnorm{A_m\cdots A_1-B_m\cdots B_1}\le\sum_{j=1}^m\opnorm{A_j-B_j}.
\label{eq:product-telescoping}
\end{equation}

To compare expectation values, let $U_1,U_2$ be unitaries, put $\Delta=U_1-U_2$ and write $\rho_i=U_i\rho U_i^\dagger$ for the output states. The identity $\rho_1-\rho_2=\Delta\rho U_1^\dagger+U_2\rho\Delta^\dagger$, together with $\norm*{ADB}_1\le\opnorm{A}\norm*{D}_1\opnorm{B}$ and $\norm*{\rho}_1=1$, implies $\norm*{\rho_1-\rho_2}_1\le2\opnorm{\Delta}$. For any observable $M$,
\begin{equation}
 \abs*{\Tr{M(\rho_1-\rho_2)}}\le\opnorm{M}\norm*{\rho_1-\rho_2}_1
 \le 2\opnorm{M}\opnorm{U_1-U_2}.
\label{eq:response-perturbation}
\end{equation}

\subsection{Cohering and imaginarity power of qubit gates}
\label{app:qubit-powers}
\begin{restatement}{Lemma}{lem:powers}
For a qubit gate $V\in\mathrm{SU}(2)$ with Bloch rotation $O=O(V)$, the cohering and imaginarity powers are
\[
 \begin{aligned}
  p_C(V)&=\sqrt{1-O_{zz}^2},\\
  p_I(V)&=\sqrt{1-O_{yy}^2}.
 \end{aligned}
\]
For coherence, the largest increase occurs for the basis state $\ket{0}$, so $p_C(V)=C(V\dyad{0}V^\dagger)$. For imaginarity, it occurs for a pure real state with Bloch vector proportional to $(O_{yx},0,O_{yz})$, provided $p_I(V)>0$.
\end{restatement}
\begin{proof}
By~\eqref{eq:bloch-measures}, each measure is the Euclidean distance of the Bloch vector from the corresponding free set,
\[
 R(\rho)=\enorm{\Pi^\perp s},\qquad\Pi^\perp=\id-\Pi,
\]
where $\Pi$ is the orthogonal projector onto the $z$ axis for coherence and onto the $xz$ plane for imaginarity.
Decomposing $s=\Pi^\perp s+\Pi s$ gives
\[
 \begin{aligned}
 R(V\rho V^\dagger)
 &=\enorm{\Pi^\perp Os}\\
 &\le\enorm{\Pi^\perp O\,\Pi^\perp s}+\enorm{\Pi^\perp O\,\Pi s}\\
 &\le R(\rho)+\opnorm{\Pi^\perp O\,\Pi},
 \end{aligned}
\]
using $\opnorm{\Pi^\perp O}\le1$ and $\enorm{\Pi s}\le1$.
Thus $p_R(V)\le\opnorm{\Pi^\perp O\,\Pi}$. Equality holds for a unit vector $s$ in the range of $\Pi$ that maximizes $\enorm{\Pi^\perp Os}$: the corresponding pure state is free, so $R(\rho)=0$ and $R(V\rho V^\dagger)=\opnorm{\Pi^\perp O\,\Pi}$.
It remains to compute this norm for each resource.
For coherence the only possibly nonzero entries of $\Pi^\perp O\,\Pi$ are $O_{xz}$ and $O_{yz}$ in the $z$ column. This column of $O$ has unit length, so the operator norm is $(O_{xz}^2+O_{yz}^2)^{1/2}=\sqrt{1-O_{zz}^2}$.
For imaginarity they are $O_{yx}$ and $O_{yz}$ in the $y$ row, which also has unit length, giving $\sqrt{1-O_{yy}^2}$.
A maximizing vector for coherence is $s=e_z$, the Bloch vector of $\ket{0}$, because the range of $\Pi$ is the $z$ axis.
For imaginarity, $\enorm{\Pi^\perp Os}=\abs*{O_{yx}s_x+O_{yz}s_z}$ for $s$ in the $xz$ plane. If $p_I(V)>0$, the Cauchy-Schwarz inequality shows that this is largest when $s$ is proportional to $(O_{yx},0,O_{yz})$.
The inverse gate has the Bloch rotation $O^{\tp}$, with the same diagonal entries and therefore the same powers.
\end{proof}

\subsection{Polynomial approximation}
\label{app:approximation-tools}
\label{app:coefficient-estimates}
A trigonometric polynomial of degree at most $L$ is a function
$p(t)=\sum_{k=-L}^Lc_ke^{\ii kt}$.
It is real-valued exactly when $c_{-k}=\overline{c_k}$ for every $k$.
The universality proof in Appendix~\ref{app:synthesis} uses Fej\'er means to approximate a target function and Fej\'er--Riesz factorization to construct the circuit amplitudes. The nonaffinity bound in Appendix~\ref{app:coherence} uses Bernstein's inequality to control derivatives and~\eqref{eq:coefficient-bessel} to control Chebyshev coefficients.

\paragraph{Fej\'er means.}
For a real continuous $2\pi$-periodic function $h$ and an integer $N\ge1$, set
\[
 \begin{aligned}
  h_N(t)&=\int_{-\pi}^{\pi}K_N(s)h(t-s)\,\frac{ds}{2\pi},\\
  K_N(s)&=\frac1N\abs{\sum_{k=0}^{N-1}e^{\ii ks}}^2.
 \end{aligned}
\]
The Fej\'er means $h_N$ are trigonometric polynomials of degree at most $N-1$ and converge uniformly to $h$~\cite[Chap.~III, Theorem~(3.4)]{Zygmund2002}.
The kernel $K_N$ is nonnegative and has mean one, so every $h_N(t)$ lies between the minimum and maximum of $h$. Since $K_N$ is even, $h_N$ is even whenever $h$ is.

\paragraph{Fej\'er--Riesz factorization.}
If a trigonometric polynomial $p$ of degree at most $L$ is nonnegative on $\mathbb R$, then
\begin{equation}
 p(t)=\abs*{P(e^{\ii t})}^2
\label{eq:fejer-riesz}
\end{equation}
for a polynomial $P$ of degree at most $L$~\cite{DritschelRovnyak2009}.
For $p\ne0$, the factor without zeros in the open unit disk and normalized by $P(0)>0$ is unique~\cite[Sec.~2]{DritschelRovnyak2009}.
If $p$ is even, conjugating the coefficients of $P$ gives a factor with the same properties.
Uniqueness therefore implies that the coefficients of $P$ are real. For $p=0$, take $P=0$.

\paragraph{Bernstein's inequality.}
A real trigonometric polynomial $p$ of degree at most $L$ satisfies~\cite[Theorem~1.1]{QueffelecZarouf2019}
\begin{equation}
 \sup_{t\in\mathbb R}\abs*{p'(t)}\le L\sup_{t\in\mathbb R}\abs*{p(t)}.
\label{eq:bernstein}
\end{equation}

\paragraph{Chebyshev coefficients.}
For a continuously differentiable function $h$ on $[-1,1]$, put $H(\theta)=h(\cos\theta)$ on $[0,\pi]$.
With the normalization in~\eqref{eq:cheb}, the Chebyshev coefficients are~\cite[Chap.~3]{Trefethen2019}
\[
 \begin{aligned}
 \beta_0&=\frac1\pi\int_0^\pi H(\theta)\,d\theta,\\
 \beta_n&=\frac2\pi\int_0^\pi H(\theta)\cos(n\theta)\,d\theta \quad(n\ge1).
 \end{aligned}
\]
For $n\ge1$, integration by parts gives
\[
 \frac2\pi\int_0^\pi H'(\theta)\sin(n\theta)\,d\theta
 =-\frac{2n}{\pi}\int_0^\pi H(\theta)\cos(n\theta)\,d\theta
 =-n\beta_n.
\]
Bessel's inequality~\cite[Theorem~8.12]{Rudin1976} bounds the sum of squared coefficients in an orthonormal system by the squared $L^2$ norm of the function. Applied on $[0,\pi]$ to the normalized cosine functions when $k=0$ and sine functions when $k=1$, it gives
\begin{equation}
 \sum_{n\ge1}n^{2k}\abs*{\beta_n}^2
 \le\frac2\pi\int_0^\pi\abs*{H^{(k)}(\theta)}^2\,d\theta
 \le2\max_{\theta\in[0,\pi]}\abs*{H^{(k)}(\theta)}^2,\qquad k=0,1.
\label{eq:coefficient-bessel}
\end{equation}

\section{Universality of the one-qubit model}
\label{app:synthesis}
We derive the uniform approximation statement used in our proofs from the polynomial characterization of circuit unitaries by Yu et al.~\cite[Lemmas~1 and~3]{Yu2022} and the approximation results in Appendix~\ref{app:approximation-tools}. The version with unrestricted gates is used in Theorem~\ref{thm:magic-mean} and for the factors of the product class in Theorem~\ref{thm:entangling-power}; the version for even targets with real gates is used in Theorem~\ref{thm:i1}.

\begin{proposition}[Uniform approximation on the input intervals]
\label{prop:one-qubit-universality}
For each encoding in Table~\ref{tab:encodings}, circuits with unrestricted trainable gates and depth, input $\rho_r$ and $Z$ readout approximate every $g\in\cont(\X;[-r,r])$ uniformly. Here $\X=[-1,1]$ for $R_z(x)$ and $R_y(x)$, and $\X=[0,1]$ for amplitude encoding. For phase encoding, real trainable gates suffice for every even target.
\end{proposition}
\begin{proof}
The model function for the input $\rho_r$ is $r$ times the one for the pure input $\ket{0}$ (Section~\ref{sec:model}), so it suffices to treat $r=1$.

We first recall which polynomials can occur as circuit amplitudes. Let $P$ and $Q$ be Laurent polynomials in $z=e^{\ii t/2}$ of degree at most $L$ in which all exponents have the parity of $L$, that is, of the form $\sum_{j=0}^Lc_jz^{2j-L}$, and let $\abs*{P(t)}^2+\abs*{Q(t)}^2=1$ for all real $t$.
By Ref.~\cite[Lemma~3]{Yu2022}, there is then a circuit~\eqref{eq:circuit} with $L$ encoding gates $R_z(t)$ that implements the unitary
\[
 U(t)=\begin{pmatrix}P(t)&-Q(t)\\\overline{Q(t)}&\overline{P(t)}\end{pmatrix},
\]
with trainable gates that are products of rotations about the $z$ and $y$ axes.
If $P$ and $Q$ have real coefficients, rotations $R_y(\theta_j)$ suffice as trainable gates~\cite[Lemma~1]{Yu2022}, and these have real matrices.
The state $U(t)\ket{0}$ has the amplitudes $P(t)$ and $\overline{Q(t)}$, so the model function of the circuit is $\abs*{P}^2-\abs*{Q}^2$.

To construct such amplitudes for a target function, let $h:J\to[-1,1]$ be continuous, with $J=[-1,1]$ or $J=[0,\pi]$. Write $J=[a,b]$. On $[b,a+2\pi]$, interpolate linearly from $h(b)$ to $h(a)$ and then extend periodically. This gives a continuous periodic extension with values in $[-1,1]$.
Its Fej\'er means are real trigonometric polynomials with values in $[-1,1]$, and they converge to it uniformly (Appendix~\ref{app:approximation-tools}).
Let $q$ be one of these means and $L$ its degree.
The polynomials $(1\pm q)/2$ are nonnegative, and the Fej\'er--Riesz factorization~\eqref{eq:fejer-riesz} gives
\[
 \frac{1+q(t)}2=\abs*{P_0(e^{\ii t})}^2,\qquad\frac{1-q(t)}2=\abs*{Q_0(e^{\ii t})}^2
\]
with polynomials $P_0$ and $Q_0$ of degree at most $L$.
Set $P(t)=z^{-L}P_0(z^2)$ and $Q(t)=z^{-L}Q_0(z^2)$ with $z=e^{\ii t/2}$.
A monomial of degree $k$ in $P_0$ or $Q_0$ acquires the exponent $2k-L$, so $P$ and $Q$ have the form required above, and $\abs*{P}^2+\abs*{Q}^2=1$ because $\abs*{z}=1$.
The corresponding circuit has the model function $\abs*{P}^2-\abs*{Q}^2=q$.
Hence model functions with phase encoding approximate $h$ uniformly on $J$.

If $h$ is even on $J=[-1,1]$, its endpoint values agree, so the extension is constant on the rest of the period and is even. Its Fej\'er means are then even as well, and $P_0$ and $Q_0$ can be chosen with real coefficients (Appendix~\ref{app:approximation-tools}). Then $P$ and $Q$ have real coefficients too, and real trainable gates suffice.

Taking $J=[-1,1]$ and $h=g$ proves the claim for phase encoding.
For $R_y$ encoding the identity $R_y(t)=R_x(-\pi/2)R_z(t)R_x(\pi/2)$ gives a fixed change of basis that can be absorbed into the trainable gates, as described after Eq.~(15) of Ref.~\cite{Yu2022}, so the model functions are the same.
Amplitude encoding applies $R_y(t)$ with $t=2\arcsin\sqrt{x}$, which runs from $0$ to $\pi$ as $x$ runs from $0$ to $1$, with continuous inverse $x=\sin^2(t/2)$. Approximating $h(t)=g(\sin^2(t/2))$ on $J=[0,\pi]$ therefore gives the same uniform error for $g$ on $[0,1]$.
\end{proof}

\section{Coherence bound}
\label{app:coherence}
\begin{restatement}{Theorem}{thm:c1}
Let $f$ be the model function of the circuit~\eqref{eq:circuit} with input state $\rho_r$, $0<r\le1$, phase encoding $R_z(x)$ on $[-1,1]$, $Z$ readout and $L\ge1$ encoding calls. If its trainable gates have cumulative cohering power $P_C$, then
\[
 \begin{aligned}
  \osc(f)&\le 2r\min\{1,P_C^2/4\},\\[2pt]
  \N(f)&\le K_L\,r\min\{1,P_C^2/4\},
 \end{aligned}
\]
where $K_L=\min\{L^2/2,\sqrt{2\pi L}\}$.
At $P_C=0$ the only model functions are $f\equiv r$ and $f\equiv-r$. For $L\ge2$ the oscillation bound is attained for every $0\le P_C\le2$.
\end{restatement}
\begin{proof}
Let $q=f/r$, the output of the same circuit with the pure input $\ket{0}$ (Section~\ref{sec:model}). We first bound its oscillation over all real inputs. This will allow us to apply Bernstein's inequality~\eqref{eq:bernstein}, which requires a bound over a whole period, when estimating the nonaffinity.

At any stage of the circuit, let $a_x$ and $b_x$ be the absolute values of the amplitudes of $\ket{0}$ and $\ket{1}$ for the input $x\in\mathbb R$, so that $a_x^2+b_x^2=1$.
Set
\[
 S=\sup_{x,y\in\mathbb R}a_xb_y.
\]
Initially $b_y=0$ for every $y$, and hence $S=0$.
The encoding gate changes only the phases of the amplitudes and leaves $S$ unchanged.
For a trainable gate $V$, let $c=\abs*{V_{00}}$ and $s=\abs*{V_{10}}$ be the absolute values of its diagonal and off-diagonal entries, so that $c^2+s^2=1$ and $p_C(V)=2cs$ (Section~\ref{sec:bloch}).
By the triangle inequality, the absolute values $a_x'$ and $b_y'$ after the gate satisfy
\begin{align*}
 a_x'b_y'&\le(ca_x+sb_x)(sa_y+cb_y)\\
 &=c^2\,a_xb_y+s^2\,a_yb_x+cs\,(a_xa_y+b_xb_y)
 \le S+cs.
\end{align*}
The last inequality uses $a_xb_y,a_yb_x\le S$, $c^2+s^2=1$ and $a_xa_y+b_xb_y\le1$, the latter by Cauchy--Schwarz.
Each trainable gate therefore raises $S$ by at most $p_C(V)/2$, and at the end of the circuit $S\le P_C/2$.
At readout $q(x)=a_x^2-b_x^2=1-2b_x^2$, so for all $x,y\in\mathbb R$
\[
 \frac{\abs*{q(x)-q(y)}}{2}
 =\abs*{b_x^2-b_y^2}
 =\abs*{(a_yb_x)^2-(a_xb_y)^2}
 \le S^2\le\frac{P_C^2}{4}.
\]
Here we used $a_x^2+b_x^2=a_y^2+b_y^2=1$ and the fact that both products are at most $S$. Thus $q$ oscillates by at most $P_C^2/2$ over $\mathbb R$. Together with $\abs*{q}\le1$, this proves the bound on $\osc(f)$ in~\eqref{eq:coherence}.

When $P_C=0$, we have $S=0$ at the end of the circuit, that is, $a_xb_y=0$ for all $x$ and $y$.
Hence either $b_y=0$ for all $y$ or $a_x=0$ for all $x$, so that $f\equiv r$ or $f\equiv-r$.
Taking all trainable gates to be the identity gives $f\equiv r$. Replacing $V_0$ by $R_y(\pi)$ gives $f\equiv-r$.

To show that the oscillation bound is sharp, let $L\ge2$ and $0\le\alpha\le\pi/2$, and take $V_0=R_y(\alpha)$, $V_L=R_y(-\alpha)R_z(\pi/2)$ and identity gates in between.
The factor $R_z(\pi/2)$ changes only the phases of the entries of $V_L$, so both gates have the cohering power $\sin\alpha$, and $P_C=2\sin\alpha$.
The circuit is $R_y(-\alpha)R_z(Lx+\pi/2)R_y(\alpha)$, and~\eqref{eq:coherence-example} gives
\[
 q(x)=\cos^2\alpha+\sin^2\alpha\cos(Lx+\pi/2)=\cos^2\alpha-\sin^2\alpha\sin(Lx).
\]
The points $x=\pm\pi/(2L)$ lie in $[-1,1]$ and give $\osc(f)=2r\sin^2\alpha=rP_C^2/2$.
For $L\ge4$, $x=0$ and $x=\pi/L$ both lie in $[-1,1]$ and attain the extrema without the phase shift. Figure~\ref{fig:bounds}(a) shows this family for $L=4$.
For $L=1$, put $u=O(V_0)e_z$ and $v=O(V_1)^{\tp}e_z$. Equation~\eqref{eq:bloch-response} gives $q(x)=u_zv_z+v_\perp^{\tp}Q_xu_\perp$. As $Q_x$ rotates $u_\perp$ in the transverse plane, the second term is a cosine of amplitude $\enorm{u_\perp}\enorm{v_\perp}=p_C(V_0)\,p_C(V_1)\le P_C^2/4$. On an interval of length two a cosine oscillates by at most $2\sin1$ times its amplitude, so the bound is strict when $P_C>0$.

We now use the oscillation bound to estimate the nonaffinity. By~\eqref{eq:bloch-response}, $q$ is a polynomial of degree at most $L$ in the entries $\cos x$ and $\pm\sin x$ of $Q_x$, and hence a real trigonometric polynomial of degree at most $L$.
Let $q_*$ be the midpoint between its maximum and minimum over $\mathbb R$, and let $\Gamma=\min\{1,P_C^2/4\}$.
By the first part, $q-q_*$ is bounded by $\Gamma$ on all of $\mathbb R$.
Applying Bernstein's inequality~\eqref{eq:bernstein} once and twice gives
\begin{equation}
 \norm*{q'}_{\infty,\mathbb R}\le L\Gamma,\qquad\norm*{q''}_{\infty,\mathbb R}\le L^2\Gamma.
\label{eq:derivative-bounds}
\end{equation}
Let $\beta_n$ be the Chebyshev coefficients of $q$ on $[-1,1]$, so that $\N(f)=r\sum_{n\ge2}\abs*{\beta_n}$.
We bound this sum in two ways. The first estimate is the better one for $L\le2$, the second for $L\ge3$.

For the first estimate, subtract $q'(0)\,x$ from $q$. This changes only $\beta_1$, which does not enter the sum, and makes the derivative vanish at $x=0$. Set $G(\theta)=q(\cos\theta)-q'(0)\cos\theta$.
Then $G'(\theta)=-\sin\theta\,[q'(\cos\theta)-q'(0)]$, and the mean value theorem with~\eqref{eq:derivative-bounds} gives $\abs*{G'(\theta)}\le L^2\Gamma\,\abs*{\sin\theta\cos\theta}$.
Since $\int_0^\pi\sin^2\theta\cos^2\theta\,d\theta=\pi/8$, \eqref{eq:coefficient-bessel} with $k=1$ yields
\[
 \sum_{n\ge2}n^2\abs*{\beta_n}^2\le\frac{L^4\Gamma^2}{4}.
\]
Applying Cauchy--Schwarz to $n\abs*{\beta_n}$ and $1/n$ gives
\[
 \Bigl(\sum_{n\ge2}\abs*{\beta_n}\Bigr)^2
 \le \Bigl(\sum_{n\ge2}n^2\abs*{\beta_n}^2\Bigr)
      \Bigl(\sum_{n\ge2}\frac{1}{n^2}\Bigr)
 \le \frac{L^4\Gamma^2}{4}.
\]
For the last inequality we used $\sum_{n\ge2}n^{-2}\le\sum_{n\ge2}\bigl(1/(n-1)-1/n\bigr)=1$. Taking square roots gives the first bound.

For the second estimate, let $H(\theta)=q(\cos\theta)-q_*$. The constant shift affects only $\beta_0$.
Then $\norm*{H}_{\infty,[0,\pi]}\le\Gamma$ and $\norm*{H'}_{\infty,[0,\pi]}\le\norm*{q'}_{\infty,\mathbb R}\le L\Gamma$. Adding~\eqref{eq:coefficient-bessel} for $k=0$ and, divided by $L^2$, for $k=1$ gives
\[
 \sum_{n\ge1}\Bigl(1+\frac{n^2}{L^2}\Bigr)\abs*{\beta_n}^2\le4\Gamma^2.
\]
The Cauchy--Schwarz inequality with these weights yields
\[
 \Bigl(\sum_{n\ge2}\abs*{\beta_n}\Bigr)^2
 \le\Bigl(\sum_{n\ge1}\frac1{1+n^2/L^2}\Bigr)\,4\Gamma^2
 \le\frac{\pi L}{2}\,4\Gamma^2,
\]
because $t\mapsto(1+t^2/L^2)^{-1}$ decreases for $t\ge0$, so its integral on $[n-1,n]$ bounds its value at $n$. The integral on $[0,\infty)$ is $\pi L/2$.
Hence $\sum_{n\ge2}\abs*{\beta_n}\le\sqrt{2\pi L}\,\Gamma$.

Taking the smaller bound gives $\N(f)\le r\Gamma\min\{L^2/2,\sqrt{2\pi L}\}=K_L\,r\min\{1,P_C^2/4\}$.
\end{proof}

\section{Imaginarity bound}
\label{app:imaginarity}
\begin{restatement}{Theorem}{thm:i1}
Let $f$ be the model function of the circuit~\eqref{eq:circuit} with input state $\rho_r$, $0<r\le1$, phase encoding $R_z(x)$ on $[-1,1]$ and $Z$ readout. At any depth, the cumulative imaginarity power $P_I$ of the trainable gates bounds the odd part of the output,
\[
 \supnorm{f_{\odd}}\le r\min\{1,P_I\}.
\]
Allowing arbitrary encoding depth, the uniform closure of the model functions with $P_I=0$ consists of all even functions in $\cont([-1,1];[-r,r])$. For every $g\in\cont([-1,1];[-r,r])$, the infimum of $\supnorm{f-g}$ over these circuits equals $\supnorm{g_{\odd}}$. For $L\ge2$ encoding calls the bound is attained for every $0\le P_I\le1$.
\end{restatement}
\begin{proof}
We compare the circuit with its complex conjugate, as outlined in Section~\ref{sec:imag}.
Complex conjugation maps $Y$ to $-Y$ and leaves $X$ and $Z$ unchanged.
On Bloch vectors it therefore acts as the reflection $T=\operatorname{diag}(1,-1,1)$, and $O(\overline V)=TO(V)T$ for every $V\in\mathrm{SU}(2)$.

For a single gate, write $O=O(V)$.
Conjugation by $T$ reverses the sign of exactly those entries of $O$ that connect the $y$ axis with the $xz$ plane, so that
\[
 O-TOT=2\begin{pmatrix}0&O_{xy}&0\\O_{yx}&0&O_{yz}\\0&O_{zy}&0\end{pmatrix}.
\]
Put $u=(O_{xy},O_{zy})^{\tp}$ and $v=(O_{yx},O_{yz})^{\tp}$. Since the $y$ column and row of $O$ have unit length, $\enorm{u}^2=\enorm{v}^2=1-O_{yy}^2$.
For a unit vector $s$, write $w=(s_x,s_z)^{\tp}$. The $xz$ component of $(O-TOT)s/2$ is $s_yu$, and its $y$ component is $v^{\tp}w$. Hence
\[
 \tfrac14\enorm{(O-TOT)s}^2
 =s_y^2\enorm{u}^2+\abs*{v^{\tp}w}^2
 \le (1-O_{yy}^2)\bigl(s_y^2+\enorm{w}^2\bigr)=1-O_{yy}^2.
\]
Equality holds for $s=e_y$. Thus Lemma~\ref{lem:powers} gives
\begin{equation}
 \opnorm{O-TOT}=2\sqrt{1-O_{yy}^2}=2p_I(V).
\label{eq:conjugate-distance}
\end{equation}

To apply this estimate to the full circuit, write $O_j=O(V_j)$ in~\eqref{eq:bloch-response}. Since $R(-x)=TR(x)T$, $T^2=\id$ and $Te_z=e_z$,
\[
 \begin{aligned}
 f(-x)&=r\,e_z^{\tp}O_LR(-x)\cdots R(-x)O_0e_z\\
      &=r\,e_z^{\tp}(TO_LT)R(x)\cdots R(x)(TO_0T)e_z.
 \end{aligned}
\]
This is the output at $x$ of the circuit with the complex-conjugated gates $\overline{V_j}$.
It differs from $f(x)$ only through the factors $TO_jT$ in place of $O_j$, so~\eqref{eq:product-telescoping} and~\eqref{eq:conjugate-distance} give
\[
 \abs*{f(x)-f(-x)}\le r\sum_{j=0}^L\opnorm{O_j-TO_jT}=2rP_I.
\]
Since $f_{\odd}(x)=[f(x)-f(-x)]/2$, this means $\abs*{f_{\odd}}\le rP_I$.
Moreover, $\abs*{f_{\odd}}\le r$ because $f$ takes values in $[-r,r]$.
This proves~\eqref{eq:imagbound}.

At zero power,~\eqref{eq:imagbound} implies that every model function is even, and so is every uniform limit of such functions.
Conversely, Proposition~\ref{prop:one-qubit-universality} gives uniform approximation of every even target in $\cont([-1,1];[-r,r])$ by circuits whose trainable gates have real matrices.
Such gates are rotations $R_y(\theta)$, for which $O_{yy}=1$ and hence $p_I=0$.

For an arbitrary target $g\in\cont([-1,1];[-r,r])$, let $f$ be a model function with $P_I=0$. Since $f$ is even, $g_{\odd}=(g-f)_{\odd}$, and therefore $\supnorm{g_{\odd}}\le\supnorm{g-f}$.
Conversely, the even part $g-g_{\odd}$ takes the value $[g(x)+g(-x)]/2$ at $x$, so it is continuous and takes values in $[-r,r]$.
It is therefore a uniform limit of model functions with $P_I=0$, and its distance from $g$ is $\supnorm{g_{\odd}}$.

To show that the bound is sharp, consider $R_x(\beta)R_z(x)^LR_y(\pi/2)$. The first rotation sends the Bloch vector to $re_x$, and the encoding gates then send it to $r(\cos Lx,\sin Lx,0)$. The final rotation gives the measured $z$ component $r\sin\beta\sin(Lx)$.
Its cumulative imaginarity power is $P_I=\abs*{\sin\beta}$, since $R_y(\pi/2)$ and the intervening identity gates are real.
For $L\ge2$ the point $x=\pi/(2L)$ lies in $[-1,1]$, and there $\abs*{f_{\odd}}=rP_I$.
\end{proof}

\section{The two-qubit model}
\label{app:entangling-response}
\subsection{Universality}
\label{app:two-qubit-universality}
\begin{proposition}[Uniform approximation with two qubits]
\label{prop:two-qubit-universality}
In the model of Section~\ref{sec:multi} with unrestricted depth and cumulative entangling power, every $g\in\cont(D_1\times D_2;[-1,1])$ can be approximated uniformly. Each qubit may use any of the three local encodings in Table~\ref{tab:encodings}.
\end{proposition}
\begin{proof}
We use a SWAP built from $R_{ZZ}$ gates to give the first qubit access to both inputs, then approximate a rotation $R_y(h)$ on this qubit.
Here $h$ is a trigonometric polynomial close to $\arccos g$, so the ideal rotation has a $Z$ expectation value $\cos h$ close to $g$.

First, we express the three encodings in terms of their rotation angles $t_i$ instead of the inputs $x_i$: $t_i=x_i\in J_i=[-1,1]$ for the two rotation encodings, and $t_i=2\arcsin\sqrt{x_i}\in J_i=[0,\pi]$ for amplitude encoding.
As in the proof of Proposition~\ref{prop:one-qubit-universality}, this change of coordinates and a fixed local change of basis, which turns $R_y(t_i)$ into $R_z(t_i)$, preserve the uniform error. We can therefore use phase encoding on $J=J_1\times J_2$, writing $g(t_1,t_2)$ for the target in these coordinates.

To implement the SWAP using the allowed gates, note that the operators $Q\otimes Q$ with $Q\in\{X,Y,Z\}$ commute, and their sum is $2\,\mathrm{SWAP}-\id$.
Exponentiating this identity gives
\begin{equation}
 \mathrm{SWAP}=e^{\ii\pi/4}\prod_{Q\in\{X,Y,Z\}}e^{-\ii\pi Q\otimes Q/4},
\label{eq:swap}
\end{equation}
and each factor is an $R_{ZZ}(\pi/2)$ gate conjugated by local gates; the powers of these gates enter $P_E$.
Conjugating the encoding gate on the second wire by this product gives $R_z(t_2)$ on the first wire and the identity on the second; the global phase cancels.
The first qubit can therefore use both encoding gates $R_z(t_1)$ and $R_z(t_2)$, interspersed with arbitrary data-independent gates.
At readout the second qubit is in $\ket0$, so the first qubit's $Z$ expectation value equals the two-qubit $Z\otimes Z$ expectation value.

We next construct circuits that uniformly approximate $R_y(a\cos\theta)$, where $\theta=k_1t_1+k_2t_2+\beta$, $k_1,k_2\in\mathbb Z$ and $a,\beta\in\mathbb R$.
The gate $W=R_z(\theta)$ is a product of encoding gates, their inverses and the fixed gate $R_z(\beta)$; the inverses are obtained by conjugation with $R_x(\pi)$.
Let $T=-\ii Y/2$, so that $R_y(a)=e^{aT}$.
Conjugation by $W$ rotates $Y$ about the $z$ axis by the angle $\theta$, and conjugation by $W^\dagger$ rotates it by $-\theta$, so that
\[
 WTW^\dagger+W^\dagger TW=2\cos\theta\,T.
\]
The Lie product formula expresses the exponential of a sum of matrices as a limit of alternating products of their exponentials~\cite[Theorem~2.11]{Hall2015}.
Applied to the two generators above, it shows that the circuits
\[
 \bigl[\,WR_y(a/(2n))W^\dagger\;W^\dagger R_y(a/(2n))W\,\bigr]^n
\]
converge to $e^{a\cos\theta\,T}=R_y(a\cos\theta)$ as $n\to\infty$, for every real $a$.
The convergence is uniform on $J$. To see this, put $u=a/(2n)$, $A=WTW^\dagger$ and $B=W^\dagger TW$. Each bracket is $e^{uA}e^{uB}$, while $e^{u(A+B)}=R_y(a\cos\theta/n)$.
Both Taylor expansions start with $\id+u(A+B)$, and the remainders are $O(u^2)$ in operator norm, uniformly in $\theta$, since $\opnorm{A}=\opnorm{B}=1/2$.
The bracket and $R_y(a\cos\theta/n)$ are unitary, so~\eqref{eq:product-telescoping} bounds the error between their $n$th powers by $n\,O(u^2)=O(a^2/n)$.

This construction extends to every real trigonometric polynomial $h(t_1,t_2)$, since $h$ is a finite sum of terms $a\cos(k_1t_1+k_2t_2+\beta)$.
The corresponding rotations about the $y$ axis commute and multiply to $R_y(h)$, so circuits on the first qubit approximate $R_y(h)$ uniformly on $J$.
To approximate an arbitrary continuous target, we use the Stone--Weierstrass theorem: a real algebra of continuous functions on a compact set is uniformly dense in the space of continuous real-valued functions if it contains the constants and separates points~\cite[Theorem~7.32]{Rudin1976}.
The trigonometric polynomials form such an algebra on $J$. Indeed, sine and cosine distinguish any two points in each coordinate interval $J_i$, since its length is less than $2\pi$.

For a continuous $g:J\to[-1,1]$ and $\varepsilon>0$, we may therefore choose a trigonometric polynomial $h$ with $\supnorm{h-\arccos g}\le\varepsilon/2$.
The gate $R_y(h)$ has the $Z$ expectation value $\cos h$ on $\ket0$, and $\abs*{\cos h-g}\le\abs*{h-\arccos g}\le\varepsilon/2$.
By~\eqref{eq:response-perturbation}, a circuit within $\varepsilon/4$ of $R_y(h)$ in operator norm changes this expectation value by at most $\varepsilon/2$.
\end{proof}

\paragraph{Relation to other universality results.}
Proposition~\ref{prop:two-qubit-universality} is not needed for the resource bounds. It establishes that the model constrained by Theorem~\ref{thm:entangling-power} is universal when the entangling power is unrestricted.
To our knowledge, universality has not been shown before for a model of this kind, with two qubits at every accuracy, each encoding its own input, $R_{ZZ}$ gates and $Z\otimes Z$ readout. The closest results concern other models.
Yu et al.~\cite[Theorem~2]{Yu2024} prove that data re-uploading circuits with fixed encoding gates, measured with $Z$ on one qubit, approximate every continuous function from $[0,1]^d$ to $[-1,1]$ uniformly.
Their construction adds the terms of a Bernstein polynomial of degree $n$ by a linear combination of unitaries with an index register and controlled gates, and its number of qubits grows as $O(d\log n)$.
P\'erez-Salinas et al.~\cite[Theorem~1]{PerezSalinas2024Minimal} study the setting to which our proof reduces once the SWAP has moved both encoding gates to the first qubit: one qubit with a separate fixed encoding gate for each input coordinate.
Their statement concerns the amplitude $\bra{0}U\ket{0}$, with the coordinate $s_k\in[0,1]$ encoded by $e^{\ii\pi s_kZ}$.
At $s_k=0$ and $s_k=1$ this gate equals $\id$ and $-\id$, so the encoding gates at the two endpoints differ only by a global phase.
Holding the other coordinates fixed, the amplitudes at the two endpoints therefore have the same modulus, which rules out uniform approximation of targets whose moduli differ there. For the target $s\mapsto s_1$ this forces a uniform error of at least $1/2$.
In our proof the output is a $Z$ expectation value, and the encoding angles range over intervals shorter than a period. The two ends of an input interval therefore need not give the same output.
Other universality results use trainable input weights~\cite{PerezSalinas2021}. Another approach allows the number of qubits to grow with the required accuracy and chooses an observable that depends on the target. Results of this kind give approximation in $L^2$~\cite[App.~C]{Schuld2021} and, for continuous targets of $N$ inputs on a domain compactly contained in $[0,2\pi]^N$, in the uniform norm~\cite[Theorem~3]{Manzano2025}. Ref.~\cite{LiuPira2026} studies the depth required to replace trainable input weights by fixed encoding gates.

\subsection{Entangling power of \texorpdfstring{$R_{ZZ}$}{RZZ}}
\label{app:rzz-power}
\begin{restatement}{Lemma}{lem:rzz-power}
For every $\gamma\in\mathbb R$, the largest increase in generalized entanglement robustness over all two-qubit inputs is
\[
 p_E\bigl(R_{ZZ}(\gamma)\bigr)=\abs*{\sin\gamma}.
\]
A product input attains the maximum.
\end{restatement}
\begin{proof}
We first express the action of $R_{ZZ}(\gamma)$ in terms of its action at three fixed angles. The resulting identity~\eqref{eq:three-angles} will also be used to bound the distance from product functions in Appendix~\ref{app:product-distance}.
Let $P=Z\otimes Z$ and $\Phi_\gamma(\rho)=R_{ZZ}(\gamma)\,\rho\,R_{ZZ}(\gamma)^\dagger$.
Since
\begin{equation}
 R_{ZZ}(\gamma)=\cos(\gamma/2)\,\id-\ii\sin(\gamma/2)\,P,
\label{eq:coupling-expansion}
\end{equation}
we have
\[
 \Phi_\gamma(\rho)=\frac{\rho+P\rho P}{2}+\cos\gamma\,\frac{\rho-P\rho P}{2}+\sin\gamma\,\frac{\ii(\rho P-P\rho)}{2}.
\]
Write $A$, $B$ and $C$ for the three linear maps of $\rho$ on the right-hand side, so that $\Phi_\gamma=A+B\cos\gamma+C\sin\gamma$. The values at three angles determine them: $\Phi_0=A+B$ and $\Phi_{\pm\pi/2}=A\pm C$ give $A=(\Phi_{\pi/2}+\Phi_{-\pi/2})/2$, $C=(\Phi_{\pi/2}-\Phi_{-\pi/2})/2$ and $B=\Phi_0-A$. Substitution gives
\begin{equation}
 \Phi_\gamma=\cos\gamma\,\Phi_0+b_+\Phi_{\pi/2}-b_-\Phi_{-\pi/2},
\label{eq:three-angles}
\end{equation}
where $b_\pm=[\sin\gamma\pm(1-\cos\gamma)]/2$.
For $0\le\gamma\le\pi/2$ both $b_+$ and $b_-$ are nonnegative, and $b_++b_-=\sin\gamma$.
It suffices to consider these angles.
Shifting $\gamma$ by $\pi$ multiplies $R_{ZZ}(\gamma)$ by $-\ii P$, and conjugation by $X\otimes\id$ reverses the sign of $\gamma$. Both $-\ii P$ and $X\otimes\id$ are local gates, which leave $R_g$ unchanged, and neither operation changes $\abs*{\sin\gamma}$.

To bound the increase in robustness, note that every two-qubit state $\omega$ satisfies $\omega\le2\tau$ with $\tau=[\omega+(Z\otimes\id)\,\omega\,(Z\otimes\id)]/2$, because $2\tau-\omega=(Z\otimes\id)\,\omega\,(Z\otimes\id)$ is positive semidefinite.
This averaging removes the off-diagonal blocks of $\omega$ in the computational basis of the first qubit. Thus $\tau$ is a mixture of two product states, one for each basis state of the first qubit, and therefore separable.
Let $\rho$ be any state, with $t=R_g(\rho)$ and $\rho\le(1+t)\sigma$ for some $\sigma\in\mathrm{Sep}$ (Section~\ref{sec:measures}).
In~\eqref{eq:three-angles}, we have $\Phi_0(\rho)=\rho\le(1+t)\sigma$ and $\Phi_{\pi/2}(\rho)\le2\tau$ for a separable state $\tau$.
The coefficients $\cos\gamma$ and $b_+$ are nonnegative, while $-b_-\Phi_{-\pi/2}(\rho)$ is negative semidefinite. Thus
\[
 \Phi_\gamma(\rho)\le(1+t)\cos\gamma\,\sigma+2b_+\tau.
\]
The coefficients sum to $(1+t)\cos\gamma+2b_+=1+t\cos\gamma+\sin\gamma$. Dividing the right-hand side by this sum gives a convex mixture of the separable states $\sigma$ and $\tau$.
Hence $R_g(\Phi_\gamma(\rho))\le t\cos\gamma+\sin\gamma\le R_g(\rho)+\sin\gamma$.

To show that this bound is attained on a product input, we use the pure-state formula for robustness.
Write the Schmidt decomposition of a normalized pure state as $\ket{\psi}=\sum_i\lambda_i\ket{u_i}\otimes\ket{v_i}$, with orthonormal families on each qubit and $\lambda_i\ge0$, $\sum_i\lambda_i^2=1$. Its robustness is $R_g(\dyad{\psi})=(\sum_i\lambda_i)^2-1$~\cite{VidalTarrach1999,Steiner2003}.
Let $\ket{\pm}=(\ket{0}\pm\ket{1})/\sqrt2$. By~\eqref{eq:coupling-expansion}, the state $R_{ZZ}(\gamma)\ket{++}=\cos(\gamma/2)\ket{++}-\ii\sin(\gamma/2)\ket{--}$ has Schmidt coefficients $\cos(\gamma/2)$ and $\sin(\gamma/2)$.
The gate thus raises the robustness of the product state $\ket{++}$ from zero to $2\cos(\gamma/2)\sin(\gamma/2)=\sin\gamma$.
\end{proof}
Since $R_{ZZ}(-\gamma)$ equals $R_{ZZ}(\gamma)$ conjugated by the local gate $X\otimes\id$, the largest possible decrease in robustness equals the largest increase. The definition based on the absolute change then gives the same power $\abs*{\sin\gamma}$.

\subsection{Distance from product functions}
\label{app:product-distance}
\begin{restatement}{Theorem}{thm:entangling-power}
Let $f$ be a model function of the two-qubit circuit of Section~\ref{sec:multi}, and let $\mathcal P$ be the products of continuous factors $a:D_1\to[-1,1]$ and $b:D_2\to[-1,1]$. At any depth,
\[
 \dist(f,\mathcal P)\le\min\{1,P_E\}.
\]
Allowing arbitrarily many local encoding calls, the uniform closure of the model functions with $P_E=0$ is exactly $\mathcal P$. To approximate $g\in\cont(D_1\times D_2;[-1,1])$ within uniform error $\varepsilon$ requires
\[
 P_E\ge\bigl(\dist(g,\mathcal P)-\varepsilon\bigr)_+.
\]
The distance bound is attained for every $P_E\in[0,1]$.
\end{restatement}
\begin{proof}
We compare the circuit with a product circuit by removing its entangling gates one at a time.
As in Appendix~\ref{app:rzz-power}, we may assume $0\le\gamma_j\le\pi/2$ for all $j$.
The local gates $-\ii Z\otimes Z$ and $X\otimes\id$ that reduce an angle to this range become part of the circuit, and they do not change $\abs*{\sin\gamma_j}$.
Fix the input and one of the gates, with angle $\gamma$, and let $F(t)$ be the output of the circuit in which this gate is replaced by $R_{ZZ}(t)$.
The output is linear in the state after the gate, so~\eqref{eq:three-angles} gives
\[
 F(\gamma)-\cos\gamma\,F(0)=b_+F(\pi/2)-b_-F(-\pi/2).
\]
All outputs lie in $[-1,1]$, and $b_++b_-=\sin\gamma$, so that $\abs*{F(\gamma)-\cos\gamma\,F(0)}\le\sin\gamma$.
Since $R_{ZZ}(0)=\id$, $F(0)$ is the output with the gate removed.

We now apply this estimate successively to $R_{ZZ}(\gamma_1),\ldots,R_{ZZ}(\gamma_m)$.
Let $f_j$ be the model function of the circuit without the first $j$ of them, and $\kappa_j=\cos\gamma_1\cdots\cos\gamma_j$, with $f_0=f$ and $\kappa_0=1$.
Since $\kappa_j=\kappa_{j-1}\cos\gamma_j$, the following sum telescopes,
\[
 f-\kappa_mf_m
 =\sum_{j=1}^m\kappa_{j-1}\bigl(f_{j-1}-\cos\gamma_j\,f_j\bigr).
\]
The one-gate estimate bounds the uniform norm of each difference in parentheses by $\abs*{\sin\gamma_j}$. Since $\abs*{\kappa_{j-1}}\le1$, the triangle inequality gives
\begin{equation}
 \supnorm{f-\kappa_mf_m}\le\sum_{j=1}^m\abs*{\sin\gamma_j}=P_E.
\label{eq:entangling-replacement}
\end{equation}
The last circuit contains only local gates. Hence $f_m$ is a product of two one-qubit model functions, and $\kappa_mf_m$ lies in $\mathcal P$ because $\abs*{\kappa_m}\le1$.
The distance from $\mathcal P$ is also at most one, since $0\in\mathcal P$ and $\abs*{f}\le1$.
This proves~\eqref{eq:entangling-bound}.

We next identify the functions approximable with zero entangling power.
At $P_E=0$, \eqref{eq:entangling-replacement} gives $f=\kappa_mf_m\in\mathcal P$.
Conversely, by Proposition~\ref{prop:one-qubit-universality} with $r=1$, each factor of a function in $\mathcal P$ is a uniform limit of one-qubit model functions, which require only local gates.
It remains to show that $\mathcal P$ is uniformly closed.
Let $h_n\in\mathcal P$ converge uniformly to $h$. Each $h_n$ is a product, so $h_n(x_1,x_2)h_n(y_1,y_2)=h_n(x_1,y_2)h_n(y_1,x_2)$ for all inputs. Passing to the limit at these four points gives the same identity for $h$, which is continuous and takes values in $[-1,1]$.
If $h\equiv0$, then $h\in\mathcal P$. Otherwise choose $(y_1,y_2)$ at which $\abs*{h}$ is maximal.
Then $a(x_1)=h(x_1,y_2)$ and $b(x_2)=h(y_1,x_2)/h(y_1,y_2)$ are continuous and bounded by one, and $h=ab$.
Hence the uniform closure at $P_E=0$ is exactly $\mathcal P$.

Finally, the necessary power bound~\eqref{eq:entangling-target} follows from~\eqref{eq:entangling-bound} and the triangle inequality $\dist(g,\mathcal P)\le\supnorm{g-f}+\dist(f,\mathcal P)$. Lemma~\ref{lem:product-sharpness} below proves the sharpness assertion.
\end{proof}

\subsection{Sharpness of the distance bound}
\label{app:product-sharpness}
\begin{lemma}[Circuits saturating the distance bound]
\label{lem:product-sharpness}
For every $p\in[0,1]$ and every choice of the two local encodings, there is a circuit with two encoding calls on each qubit and one $R_{ZZ}$ gate whose model function satisfies $\dist(f,\mathcal P)=P_E=p$.
\end{lemma}
\begin{proof}

Each qubit applies its encoding gate twice.
With the encoding angles $t_i$ of Appendix~\ref{app:two-qubit-universality} and suitable local gates, qubit $i$ is then rotated by $R_y(\theta_i)$, where $\theta_i=2t_i$ for the two rotation encodings and $\theta_i=2t_i-\pi$ for amplitude encoding.
Indeed, conjugation by $R_x(-\pi/2)$ turns $R_z(t)$ into $R_y(t)$, and the gate $R_y(-\pi)$ shifts the angle.
In every case $\theta_i$ is a continuous increasing function of $x_i$ whose range contains $[-\pi/2,\pi/2]$.

Let $G_\gamma=e^{-\ii\gamma X\otimes Y/2}$ with $0\le\gamma\le\pi/2$, which is $R_{ZZ}(\gamma)$ conjugated by local gates, and apply $[R_y(\theta_1)\otimes R_y(\theta_2)]\,G_\gamma$ to $\ket{00}$.
This circuit has $P_E=\sin\gamma$.
Since $(X\otimes Y)\ket{00}=\ii\ket{11}$,
\[
 G_\gamma\ket{00}=\cos(\gamma/2)\ket{00}+\sin(\gamma/2)\ket{11}.
\]
In this state $\expval{Z\otimes Z}=1$, $\expval{X\otimes X}=\sin\gamma$ and $\expval{X\otimes Z}=\expval{Z\otimes X}=0$.
With $R_y(\theta)^\dagger ZR_y(\theta)=Z\cos\theta-X\sin\theta$, the model function is
\begin{equation}
 f_\gamma=\cos\theta_1\cos\theta_2+\sin\gamma\,\sin\theta_1\sin\theta_2.
\label{eq:entangling-response}
\end{equation}

We show that $\dist(f_\gamma,\mathcal P)=p$ with $p=\sin\gamma$, and may assume $p>0$.
The product function $\cos\theta_1\cos\theta_2$ differs from $f_\gamma$ by at most $p$.
Conversely, suppose that $h(x_1,x_2)=a(x_1)b(x_2)$ in $\mathcal P$ satisfies $\supnorm{f_\gamma-h}<p$.
Where $\theta_1=\theta_2=\theta$, we have $f_\gamma=p+(1-p)\cos^2\theta\ge p$ and therefore $h>0$.
Regarded as functions of the angles, $a$ and $b$ thus satisfy $a(\theta)b(\theta)>0$ for all $\theta\in[-\pi/2,\pi/2]$.
Being continuous, they have no zeros on this interval and the same constant sign, so that $h>0$ at $\theta_1=\pi/2$, $\theta_2=-\pi/2$.
But $f_\gamma=-p$ at this point, so $h$ differs from $f_\gamma$ by more than $p$ there.
This contradiction shows that no function in $\mathcal P$ is closer to $f_\gamma$ than $p$, and therefore
\[
 \dist(f_\gamma,\mathcal P)=\sin\gamma=P_E.
\]
Choosing $\gamma=\arcsin p$ realizes each $p\in[0,1]$.
\end{proof}
An $R_{ZZ}(\gamma)$ gate placed immediately before readout can contribute to $P_E$ without changing $f$, since it commutes with $Z\otimes Z$.

\subsection{Dependence on the gate set}
\label{app:power-dependence}
The SWAP gate maps the set of separable states onto itself and therefore has zero entangling power.
If SWAP were allowed as an elementary gate, the circuit
\[
 (\id\otimes R_y(x_2))\,\mathrm{SWAP}\,(R_y(x_1)\otimes\id)
\]
would have $P_E=0$ and the model function $\cos(x_1+x_2)$, which is not a product function.
In our gate set the decomposition~\eqref{eq:swap} uses three $R_{ZZ}(\pi/2)$ gates, each of power one. This implementation of SWAP contributes three to $P_E$.

For every positive integer $n$, $R_{ZZ}(\gamma)=[R_{ZZ}(\gamma/n)]^n$.
The power $\abs*{\sin(\gamma/n)}$ of each factor tends to zero as $n\to\infty$, while the sum of the $n$ powers tends to $\abs*{\gamma}$.
Since $\abs*{\sin\gamma}\le n\abs*{\sin(\gamma/n)}$, splitting the gate never lowers the cumulative power, and in general it raises it.

\section{Approximation with Clifford trainable gates}
\label{app:magic}

The gates of zero magic power are precisely the Clifford gates (Section~\ref{sec:bloch}).
To prove Theorem~\ref{thm:magic-mean}, we first construct a quarter-turn from encoding gates and Clifford gates. Its axis makes an angle arbitrarily close to $\pi/4$ with the $z$ axis, uniformly on $\delta\le\abs*{x}\le1$ (Lemma~\ref{lem:magic-rotation}).
We then replace each trainable gate by a product of this quarter-turn, the Clifford gate $R_z(\pi/2)$ and their inverses.

\paragraph{Gate sequences.}
A finite product of Clifford gates and encoding gates $R_z(x)$ is a \emph{word} in these gates. Its encoding depth counts only the occurrences of $R_z(x)$.
Merging adjacent Clifford gates turns a word into a circuit~\eqref{eq:circuit} with Clifford trainable gates.
The inverse of a word is a word, since $R_z(-x)=XR_z(x)X$, and $R_y(x)$ is a word, because it equals $R_z(x)$ conjugated by the Clifford gate $R_x(-\pi/2)$.

\paragraph{Rotations by $\pi/2$.}
For a unit vector $n\in\mathbb R^3$ we write
\[
 S_n=e^{-\ii\pi\,n\cdot\sigma/4}=\frac{\id-\ii\,n\cdot\sigma}{\sqrt2}.
\]
On Bloch vectors this gate acts as the quarter-turn about $n$, that is, as the rotation $v\mapsto(n\cdot v)n+n\times v$ by the angle $\pi/2$.
Its square $S_n^2=-\ii\,n\cdot\sigma$ acts as the half-turn $v\mapsto2(n\cdot v)n-v$.
The gate $S=R_z(\pi/2)$ is the quarter-turn $S_{e_z}$, and conjugation moves the axis: $US_nU^\dagger=S_{O(U)n}$ for every $U\in\mathrm{SU}(2)$.
Since $\opnorm{(n-n')\cdot\sigma}=\enorm{n-n'}$, two quarter-turns satisfy
\begin{equation}
 \opnorm{S_n-S_{n'}}=\frac{\enorm{n-n'}}{\sqrt2}.
\label{eq:quarter-turn-distance}
\end{equation}

The non-Clifford rotation we approximate is the quarter-turn $S_{n_*}$ about
\[
 n_*=\frac{e_x+e_z}{\sqrt2}.
\]
Since $n_*\cdot\sigma=(X+Z)/\sqrt2$ is the Hadamard gate $H$, the square of the rotation is $S_{n_*}^2=-\ii H$.
The gate $S_{n_*}$ is not Clifford, because it maps $e_z$ to $(1/2,-1/\sqrt2,1/2)$.
Conjugating $S$ by a word $U$ produces the axis $n=O(U)e_z$. For $n$ to make approximately the same angle with the $z$ axis as $n_*$, its $z$ component $n_z=O(U)_{zz}$ must approximate $t_*=(n_*)_z=1/\sqrt2$.
Exact equality on an interval is impossible. The entry $O(U(x))_{zz}$ is a trigonometric polynomial in $x$, so equality with $t_*$ throughout an interval would make it constant on $\mathbb R$. At $x=0$, however, the word is a Clifford gate, for which the entry is $0$ or $\pm1$.

\begin{lemma}[Approximating the $z$ component of the rotation axis]
\label{lem:magic-rotation}
For all $\delta,\eta\in(0,1)$ there is a word $U$ with at most $\delta^{-4}\eta^{-2}$ encoding gates such that
\[
 \abs*{O(U(x))_{zz}-t_*}\le\eta
 \qquad\text{whenever }\delta\le\abs*{x}\le1.
\]
\end{lemma}
\begin{proof}
We iterate a map on words that drives $O(U)_{zz}$ towards $t_*$.
For $U\in\mathrm{SU}(2)$ let $n=O(U)e_z$ and $t=n_z=O(U)_{zz}$, so that $USU^\dagger=S_n$, and define
\[
 \Psi(U)=S_n\,S\,S_n^2=\bigl(USU^\dagger\bigr)\,S\,\bigl(US^2U^\dagger\bigr).
\]
If $U$ is a word with $m$ encoding gates, then $\Psi(U)$ is a word with $4m$ encoding gates.

To find the induced recursion for $t$, we compute $O(\Psi(U))_{zz}$ by applying the half-turn about $n$, the quarter-turn about $e_z$ and the quarter-turn about $n$ to $e_z$ in turn.
The half-turn gives $2tn-e_z$, and the quarter-turn about $e_z$ maps this vector to
\[
 v=(2t^2-1)e_z+2t\,e_z\times n.
\]
Since $n\cdot(e_z\times n)=0$ and $\enorm{e_z\times n}^2=1-t^2$, we obtain $n\cdot v=t(2t^2-1)$ and $e_z\cdot(n\times v)=(e_z\times n)\cdot v=2t(1-t^2)$.
The last quarter-turn maps $v$ to $(n\cdot v)n+n\times v$. Its $z$ component is $O(\Psi(U))_{zz}=F(t)$, where
\begin{equation}
 \begin{aligned}
  F(t)&=t^2(2t^2-1)+2t(1-t^2)\\
      &=t+t(1-t)(1-2t^2).
 \end{aligned}
\label{eq:magic-recursion}
\end{equation}

To estimate convergence to $t_*$, set $\rho(t)=2t(1-t)(t+t_*)$. Using $1-2t^2=-2(t-t_*)(t+t_*)$ gives the identities
\begin{equation}
 \begin{aligned}
  F(t)-t_*&=\bigl(1-\rho(t)\bigr)(t-t_*),\\
  1-F(t)&=(1-t)(1-t+2t^3).
 \end{aligned}
\label{eq:magic-contraction}
\end{equation}
For $0<t<1$ we have $0<\rho(t)<(1+t_*)/2<1$, so each iteration moves $t$ towards $t_*$ without overshooting.
Near $t=1$ this contraction is slow, because the factor $1-\rho(t)$ tends to one. There, however, $(1-F(t))/(1-t)$ tends to two, so a step nearly doubles the distance from one.
We therefore measure the error relative to this distance, by $\lambda(t)=(t-t_*)/(1-t)$, and show that each step at least halves it.
By~\eqref{eq:magic-contraction}, $\lambda(F(t))=\lambda(t)\,(1-\rho(t))/(1-t+2t^3)$, and
\[
 (1-t+2t^3)-2\bigl(1-\rho(t)\bigr)=(1-t)\bigl(2t^2+(4t_*-2)t-1\bigr).
\]
The quadratic factor increases with $t$ and is positive at $t=0.54$. For $0.54\le t<1$ the denominator $1-t+2t^3$ is therefore at least twice the numerator $1-\rho(t)$, so that
\begin{equation}
 \abs*{\lambda(F(t))}\le\frac12\,\abs*{\lambda(t)}\qquad\text{for }0.54\le t<1.
\label{eq:magic-halving}
\end{equation}

We now choose the initial word and bound the number of iterations needed for accuracy $\eta$.
Let $U_0(x)=R_y(x)$ and $U_{k+1}=\Psi(U_k)$.
Then $U_k$ is a word with $4^k$ encoding gates, and by~\eqref{eq:magic-recursion} the numbers $t_k=O(U_k(x))_{zz}$ satisfy $t_0=\cos x$ and $t_{k+1}=F(t_k)$.
For $\delta\le\abs*{x}\le1$ we have $t_0\ge\cos1>0.54$. Concavity of sine on $[0,1/2]$ gives $\sin(\abs*{x}/2)\ge\abs*{x}\sin(1/2)$, so $1-t_0=2\sin^2(x/2)\ge2\delta^2\sin^2(1/2)\ge2\delta^2/5$.
All $t_k$ lie between $t_0$ and $t_*$, so~\eqref{eq:magic-halving} applies in every step and gives
\[
 \abs*{t_k-t_*}=(1-t_k)\,\abs*{\lambda(t_k)}
 \le2^{-k}\,\frac{(1-t_k)\,\abs*{t_0-t_*}}{1-t_0}
 \le\frac{2^{-k}}{2\delta^2},
\]
because $1-t_k<1/2$ and $\abs*{t_0-t_*}<1/3$.
Let $k\ge0$ be the smallest integer with $2^k\ge1/(2\delta^2\eta)$.
Then $\abs*{t_k-t_*}\le\eta$ and $4^k\le\delta^{-4}\eta^{-2}$, so $U=U_k$ has the stated properties.
\end{proof}

\begin{restatement}{Theorem}{thm:magic-mean}
Let $g\in\cont([-1,1];[-r,r])$, $0<r\le1$, and $\varepsilon,\delta\in(0,1)$. There is a one-qubit circuit with input $\rho_r$, phase encoding $R_z(x)$, Clifford trainable gates and $Z$ readout whose model function satisfies
\[
 |f(x)-g(x)|\le\varepsilon\qquad\text{for }\delta\le |x|\le1.
\]
Its cumulative magic power is zero. If $g$ itself is realized by a circuit of the same model with unrestricted trainable gates and $D$ encoding calls, the Clifford circuit can be chosen with encoding depth polynomial in $D$, $1/\varepsilon$ and $1/\delta$.
\end{restatement}
\begin{proof}
Let $g$ be the model function of a circuit with $D$ encoding calls and trainable gates $V_0,\ldots,V_D$, and put $\eta=\varepsilon/(4(D+1))$.

We first show that $S$ and $S_{n_*}$ generate a dense subgroup of $\mathrm{SU}(2)$, so that products of these gates and their inverses can approximate each $V_j$.
Indeed, $SS_{n_*}^\dagger=\tfrac12(\id-\ii Z)(\id+\ii H)$ has the trace $1+1/\sqrt2$.
Its rotation angle $\theta$ therefore satisfies $\cos(\theta/2)=\cos^2(\pi/8)$. Boykin et al.\ prove that $\theta/\pi$ is irrational for this value~\cite{Boykin1999}. Hence the powers of $SS_{n_*}^\dagger$ are dense in the rotations about its axis.
Since $S$ is diagonal and $S_{n_*}^\dagger=(\id+\ii H)/\sqrt2$ has a nonzero off-diagonal entry, $SS_{n_*}^\dagger$ cannot be a rotation about $z$. Conjugating by the quarter-turn $S$ therefore gives rotations about a second, nonparallel axis. Rotations about these two axes generate all rotations, as in the universality proof for the Hadamard and $\pi/8$ gates~\cite[Sec.~4.5.3]{NielsenChuang2010}.
The Bloch rotations of the closure of the generated group are therefore all of $\mathrm{SO}(3)$. Every rotation is the Bloch rotation of exactly two elements $\pm V$ of $\mathrm{SU}(2)$, and since $S^4=-\id$ belongs to the group, the closure contains both. Hence it equals $\mathrm{SU}(2)$.
For a finite gate set that is closed under inverses and generates a dense subgroup of $\mathrm{SU}(2)$, the Solovay--Kitaev theorem bounds the number of gates needed for accuracy $\eta$ by a polynomial in $\log(1/\eta)$~\cite[Appendix~3]{NielsenChuang2010}.
Applied to $\{S^{\pm1},S_{n_*}^{\pm1}\}$, it gives, for every $j$, a product $w_j(S,S_{n_*})$ of at most $N$ factors within $\eta$ of $V_j$ in operator norm, with $N$ bounded by such a polynomial.
We write $w_j(S,S_n)$ for the same product with $S_{n_*}$ replaced by another quarter-turn $S_n$.

To obtain a circuit using only the allowed gates, we replace $S_{n_*}$ by a quarter-turn $USU^\dagger$ constructed from a word $U$.
Let $U$ be the word of Lemma~\ref{lem:magic-rotation} with accuracy $\eta/(2N)$.
Fix $x$ with $\delta\le\abs*{x}\le1$ and let $n=O(U(x))e_z$, so that $U(x)SU(x)^\dagger=S_n$.
Let $\phi$ be the azimuthal angle of $n$, measured from the positive $x$ axis in the $xy$ plane, and let $n'=O(R_z(\phi))n_*$.
This unit vector has the azimuth $\phi$ and the $z$ component $t_*$.
The transverse parts of $n$ and $n'$ are parallel, with the lengths $(1-n_z^2)^{1/2}$ and $t_*$.
Since $t_*=1/\sqrt2$ and $\abs*{n_z}\le1$, their difference satisfies
\[
 \abs*{\sqrt{1-n_z^2}-t_*}
 =\frac{\abs*{n_z-t_*}\,\abs*{n_z+t_*}}{\sqrt{1-n_z^2}+t_*}
 \le (1+\sqrt2)\abs*{n_z-t_*}.
\]
Adding the squares of this difference and of the difference $n_z-t_*$ of the $z$ components gives $\enorm{n-n'}\le2\sqrt2\,\abs*{n_z-t_*}$. Thus~\eqref{eq:quarter-turn-distance} gives
\[
 \opnorm{S_n-S_{n'}}\le2\,\abs*{n_z-t_*}\le\frac{\eta}{N}.
\]
The same bound holds for the inverse gates because taking adjoints preserves the operator norm. At most $N$ factors differ between $w_j(S,S_n)$ and $w_j(S,S_{n'})$, so~\eqref{eq:product-telescoping} bounds their difference by $N(\eta/N)=\eta$.
Now $S_{n'}=R_z(\phi)S_{n_*}R_z(-\phi)$, and $S$ commutes with $R_z(\phi)$, so that $w_j(S,S_{n'})=R_z(\phi)\,w_j(S,S_{n_*})\,R_z(-\phi)$.
This operator is within $\eta$ of $R_z(\phi)V_jR_z(-\phi)$.
The words $W_j=w_j(S,USU^\dagger)$ thus satisfy
\begin{equation}
 \opnorm{W_j(x)-R_z(\phi)\,V_j\,R_z(-\phi)}\le2\eta
 \qquad\text{for }j=0,\ldots,D,
\label{eq:common-azimuth}
\end{equation}
with the same angle $\phi=\phi(x)$ for every gate.

Because the same angle $\phi$ occurs for every gate, the intervening rotations cancel in the full circuit.
Replace each trainable gate $V_j$ by the word $W_j$.
Since $R_z(\phi)$ commutes with the encoding gate,
\[
 \begin{aligned}
 &R_z(\phi)V_DR_z(-\phi)\,R_z(x)\cdots R_z(x)\,R_z(\phi)V_0R_z(-\phi)\\
 &\qquad=R_z(\phi)\bigl[V_DR_z(x)\cdots R_z(x)V_0\bigr]R_z(-\phi),
 \end{aligned}
\]
and by~\eqref{eq:common-azimuth} and~\eqref{eq:product-telescoping} the unitary of the new circuit is within $2(D+1)\eta$ of this operator.
The rotation $R_z(\phi)$ commutes with the input state $\rho_r$ and with the readout $Z$, so the operator on the right has the output $g(x)$.
By~\eqref{eq:response-perturbation}, the output of the new circuit therefore differs from $g(x)$ by at most $4(D+1)\eta=\varepsilon$.

It remains to bound the encoding depth. The new circuit has only Clifford trainable gates, so $P_M=0$.
Each word $W_j$ contains at most $N$ factors $(USU^\dagger)^{\pm1}$, each with at most $2\delta^{-4}(2N/\eta)^2$ encoding gates.
In total the circuit has at most $D+8(D+1)N^3\delta^{-4}\eta^{-2}$ encoding calls.
This is at most $(D+1)^3\delta^{-4}\varepsilon^{-2}$ times a polynomial in $\log((D+1)/\varepsilon)$, and hence polynomial in $D$, $1/\varepsilon$ and $1/\delta$.

For a general $g\in\cont([-1,1];[-r,r])$, Proposition~\ref{prop:one-qubit-universality} gives a circuit with unrestricted trainable gates whose model function is within $\varepsilon/2$ of $g$.
Applying the previous steps to this function with $\varepsilon/2$ in place of $\varepsilon$ proves~\eqref{eq:magic-mean}.
\end{proof}

\paragraph{Relation to fixed-point search.}
The map $\Psi$ alternates copies of $U$ and $U^\dagger$ with Clifford phase shifts, like the recursion $U\mapsto UCU^\dagger CU$ of fixed-point quantum search.
For $C=R_z(-\pi/2)$ that recursion acts as $e\mapsto e(1-2e)^2$ on $e=(1-O(U)_{zz})/2$, as in fixed-point search with both phase shifts equal to $\pi/2$~\cite[Eq.~(5)]{Li2007FixedPoint}.
The fixed point $e=0$ is neutral, that is, the map has derivative one there. Near zero, $e$ then decays only as $1/(4n)$ after $n$ steps, while each step triples the number of copies of the preceding word. The gate count is then superpolynomial in $1/e$. For $F$, inequality~\eqref{eq:magic-halving} makes the rescaled error decrease geometrically.

\bibliographystyle{alpha}
\bibliography{references}

\end{document}